# Riemannian Trust Region Method for Minimization of Fourth Central Moment for Localized Molecular Orbitals

Aliakbar Sepehri,[1] Run R. Li,[2] Mark R. Hoffmann[1,*]

[1] *Chemistry Department, University of North Dakota, Grand Forks, North Dakota 58202-9024, USA*

[2] *Department of Chemistry and Biochemistry, Florida State University, Tallahassee, FL 32306-4390*

[*] Email: mark.hoffmann@und.edu

## Abstract

The importance of Localized Molecular Orbitals in correlation treatments beyond mean-field calculation and in the illustration of chemical bonding (and antibonding) can hardly be overstated. However, generation of orthonormal localized occupied MOs is significantly more straightforward than obtaining orthonormal localized virtual MOs. Orthonormal molecular orbitals allow facile use of highly efficient group theoretical methods (e.g., graphical unitary group approach) for calculation of Hamiltonian matrix elements in multireference configuration interaction calculations (such as MRCISD) and in quasidegenerate perturbation treatments, such as the Generalized Van Vleck Perturbation Theory. Moreover, localized MOs can elucidate qualitative understanding of bonding in molecules, in addition to high accuracy quantitative descriptions. We adopt the powers of the fourth moment cost function introduced by Jørgensen and coworkers. Because the fourth moment cost functions are prone to having multiple negative Hessian eigenvalues when starting from easily available canonical (or near canonical) MOs, standard optimization algorithms can fail to obtain the orbitals of the virtual or partially occupied spaces. To overcome this drawback, we applied a Trust Region algorithm on an orthonormal Riemannian manifold with an approximate retraction from the tangent space built into first and second derivatives of the cost function. Moreover, the Riemannian Trust Region outer iterations were coupled to truncated conjugate gradient inner loops, which avoided any costly solutions of simultaneous linear equations or eigenvector/eigenvalue solutions. Numerical examples are provided on model systems, including the high connectivity $H_{10}$ set in 1-, 2- and 3-dimensional arrangements, and on a chemically realistic description of cyclobutadiene (*c*-C$_4$H$_4$), and the propargyl radical (C$_3$H$_3$). In addition to demonstrating the algorithm on occupied and virtual blocks of orbitals, the method is also shown to work on the active space at the MCSCF level of theory.

## 1. Introduction

Many chemical processes occur in specific spatial regions of molecular structures, and the concept of local molecular orbitals (LMOs) is useful to explain these local phenomena. LMOs are one-electron functions that are limited to a fairly small volume, hence they can clearly present the atomic connections and moreover they often are approximately constant between similar units in different molecular structures.[1] The LMOs idea can be seen as a bridge between the most common paradigm in theoretical chemistry (i.e. Linear Combination of Atomic Orbitals (LCAO)-MO) and chemical intuition, and it can reduce the computational cost to describe electron correlation in many-body method.[2] The electron density is determined by the occupied molecular orbitals and their locality therefore provides additional physical insight into the molecular electronic structure. By contrast, local virtual orbitals provide less physical insight into the electronic structure. Nevertheless, the local virtual orbitals are vital for developing the local correlation methods.[3] Local support of electron correlation can help elucidate correlation effects because the short-ranged electron-electron interactions result in Coulomb holes in the wave function, whereas dispersion effects are induced by longer ranged interactions.[4] Localization has led to significant advances in the formulation of efficient many-body methods, and in particular second-order Møller-Plesset perturbation theories and coupled cluster.[3,5] For example, in coupled cluster singles and doubles (CCSD), amplitudes of the form $t_{ij}^{ab}$ is computed, where $i$, $j$ are orthonormal occupied orbitals and $a$, $b$ are orthonormal virtual orbitals. Having strongly localized virtual orbitals is essential to realizing the full benefit of localized occupied orbitals and results in sparsity of the $t$-vector and accelerates the computation because of using simple equations for orthonormal orbitals.[6] While KS-DFT below the double hybrid level is not affected by virtual orbitals, TD-DFT calculations are also more efficient with both localized occupied and unoccupied orbitals.[7]

HF or KS-DFT LMOs are obtained from canonical molecular orbitals (CMOs) by a continuous transformation. The Edmiston-Ruedenberg (ER),[8] Foster-Boys (FB),[9] and Pipek-Mezey (PM)[10] schemes are the most popular localization methods, although the Fourth Moment approaches, introduced by Jørgensen and coworkers,[4,11,12] has some desirable features, and the present work explores a new variant of FM. Edmiston and Ruedenberg defined an approach which maximizes the sum of the orbital self-repulsion energies to determine those energy localized orbitals.[8] The task of FB technique is to maximize the distance between the orbital centroids, and consequently minimizing the orbital spatial extent.[9] In the PM method, the number of atomic sites are reduced for a bonding orbitals where the sum of the squares of the atomic Mulliken population of the MOs are maximized.[10] PM orbitals are considered more chemical than the FB and ER orbitals, because they preserve $\sigma - \pi$ separation. On the other hand, determining FB and PM orbitals is significantly less computationally intensive than obtaining ER orbitals.[6] However, Second and Fourth central Moment (SM and FM) approaches are well-defined in terms of statistical measures leading to strict constraints on the orbital spatial extent; consequently, these localization functions are less system dependent than the ER and PM methods, and can be further refined using a penalty to improve their locality.[4,11] Note that the SM minimization is equivalent to the FB approach.

Traditional optimizations of orbital locality use Jacobi sweeps, which are effective for occupied orbitals because of strong and isolated minima. However, they cannot localize the more complicated virtual space in general.[11,12] Due to shallow and not well-separated local minima in the virtual space, the convergence for optimization of localization functions is slow and painful, if even possible.[4,6] Naive implementations of more modern optimization techniques [e.g., conjugate gradient and Broyden-Fletcher-Goldfarb-Shanno (BFGS) algorithm] also struggle, unless a favorable initial guess is provided.[6] Høyvik and Jørgensen[3] demonstrated that localization of Hartree-Fock orbitals can be obtained, for both occupied and virtual orbitals, with the trust-region minimization algorithm of Fletcher.[13,14] The basic concept of the trust-region method is to expand the localization function up to a second-order Taylor series and use a so-called *trust* region, in which the localization function is approximated sufficiently by the second-order Taylor expansion. The function is minimized within a given trust region and the trust region is updated in each iteration.[3] However, classic trust region approaches rely on a detailed knowledge of the eigenspectrum of the second-derivative (i.e. Hessian) matrix, which can be problematic if the Sylvester signature is not simple.

This paper is divided into four additional sections, and an Appendix. In Section 2, an explicit gradient of the fourth moment localization method is derived. In Section 3, a trust-region algorithm for Riemannian embedded manifolds is presented. In Section 4, examples of the approach for both Hartree-Fock and MCSCF variable occupancy orbitals sets are presented for a set of $H_{10}$ models, $c$-$C_4H_4$, and the propargyl radical. A summary of the work is presented in the last section, with detailed formulas for the gradients and Hessians of the objective function are given in the Appendix.

## 2. Fourth Moment (FM) Localization

A review of developments regarding to the spatially localized orthogonal occupied and virtual Hartree-Fock orbitals has recently been presented by Høyvik and Jørgensen.[3] They showed that different localization functions can produce Hartree-Fock occupied orbitals of similar locality. However, localizing virtual orbitals is more difficult than localizing occupied orbitals, and the fourth central moment cost function together with a trust radius algorithm is more robust than other schemes. For fourth moment localization, we follow Høyvik and Jørgensen[3] and consider a cost function which is the sum of the fourth central moments of the orbitals,[11]

$$\xi_{FM}^m = \sum_p (\mu_4^p)^m \qquad [1]$$

where the fourth central moment of orbital $p$ is defined

$$\mu_4^p = \langle p | v_x^4 + 2v_x^2 v_y^2 + 2v_x^2 v_z^2 + v_y^4 + 2v_y^2 v_z^2 + v_z^4 | p \rangle \qquad [2]$$

and e.g. $v_x^2 = (x - \langle p|x|p \rangle)^2$ is the variance *operator* of orbital $p$ with respect to the $x$-direction, $v_x^4 = (x - \langle p|x|p \rangle)^4$ is the corresponding kurtosis *operator*, and e.g. $v_x^2 v_y^2 = (x - \langle p|x|p \rangle)^2 (y - \langle p|y|p \rangle)^2$ is the symmetric cokurtosis *operator* in the $x$ and $y$ directions. Minimizing $\xi_{FM}^m$ while retaining the orthonormality of the orbitals generates the set of orbitals which on average are most localized for the manifold spanned by the orbitals. It should be appreciated that optimizing the orbitals is highly nonlinear and will result in an iterative algorithm since the function values appear in the operators. $m$ is an integer number which exacts a penalty on the least local orbitals. As noted by Høyvik and Jørgensen, powers of $\mu_4^p$ localize a function value more than $\mu_4^p$ itself.[3] The tail decay for occupied and virtual Hartree-Fock orbitals occurs quickly already with power two.[11] However, we will consider power one through four to evaluate the performance of the algorithm presented in this work.

Matrix optimization literature in the past few decades has seen increasing use of the framework of Riemannian manifolds, based on the seminal work of Smith.[15,16] Among other advantages, the tangent vector space can be better behaved (in terms of properties such differentiation) than the original manifold. Of course, concepts from differential geometry, such as retractions, need to be addressed. The monograph by Absil *et al.*[17] provides an algorithmic approach that is translatable into a variety of specific domain problems. Their trust-region method on Riemannian manifolds, with truncated Conjugate Gradient subproblem, method forms the core of the framework used in this article.

Hamiltonians, including Fockians, define invariant subsets of orbitals, which was exploited in the approach to macroconfigurations used heavily in earlier works from the North Dakota group.[18] The present work uses the invariance of subsets to define a particular basis, i.e. localization, for the subset. In previous work, the elevation of an orbital basis to a manifold was not exploited. Let us denote a point on this orthogonal manifold, $\mathcal{M}$, by the $n \times n$ matrix **Y**, where $n$ is the number of rotating orbitals. Technically, **Y** is a point on a chart of $\mathcal{M}$ and should carry an extra label and be denoted as e.g., $\mathbf{Y}(\vec{s})$, where $s$ represents the $n$-dimensional rotation, but we will use the less cluttered notation if there is no danger of misinterpretation. Consequently, orbitals at the origin of a chart can be written as

$$|p\rangle = \sum_{\mu}^{dim(\mu)} |\chi_\mu\rangle C_{\mu p} = \sum_{\mu}^{dim(\mu)} |\chi_\mu\rangle \sum_{q}^{n} C_{\mu q} \mathbf{Y}(0)_{qp} \qquad [3]$$

where $|\chi_\mu\rangle$ represents an atomic orbital, $dim(\mu)$ is the number of atomic orbitals, $C_{\mu p}$ is an element of the molecular coefficient matrix defining the origin of the chart, and $\mathbf{Y}(0)$ is $I_{n \times n}$, the $n$-dimensional identity matrix. Optimization on the chart is parametrized on a direction in the tangent space,

$$|\tilde{p}\rangle = |\vec{\kappa}\rangle = \sum_{q} |q\rangle [\mathbf{Y}(\vec{\kappa})]_{q\tilde{p}} \qquad [4]$$

where $|\vec{\kappa}\rangle$ contains the unique elements of the matrix operator $\hat{\kappa}$. The tangent space of an orthonormal matrix manifold is the space of skew symmetric matrices of the same dimension, $\kappa_{ij} = -\kappa_{ji}$, and a curve in the orthogonal manifold is produced on retraction from the tangent manifold, $R_Y: T_Y\mathcal{M} \to \mathcal{M}$. Although a number of retractions are possible, e.g., projective, we use the exponential map,

$$[\mathbf{Y}(\vec{\kappa})]_{q\tilde{p}} = [\exp(-\hat{\kappa})]_{q\tilde{p}}. \qquad [5]$$

Whereas definitions of Hessians on general manifolds are problematic,[17] and definitions involve all vector bundles $\mathfrak{X}(\mathcal{M})$, it is the case that Riemannian connections leave the Riemannian metric invariant and allow a local Euclidean second order approximation,

$$\hat{Q}_{\mathbf{Y}(0)}(\eta) = f(0) + \langle \operatorname{grad} f(0), \eta \rangle + \frac{1}{2}\langle \operatorname{Hess} f(0)[\eta], \eta \rangle \qquad [6]$$

where $\eta$ is a vector of unique elements of $\boldsymbol{\eta}$ on the tangent bundle, and $\operatorname{grad} f(0)$ and $\operatorname{Hess} f(0)[\eta]$ are the corresponding unique elements of the gradient and product of Hessian with vector, and, in a slight misuse of notation, $\hat{Q}_{\mathbf{Y}(0)}(\eta)$ represents one of any possible local Euclidean second order approximation of the objective function.

The explicit form of the operator generating the curve on which a local minimum is sought would be

$$\hat{\kappa} = \sum_{i>j} \kappa_{ij} \left( \hat{a}_i^\dagger \hat{a}_j - \hat{a}_j^\dagger \hat{a}_i \right) \qquad [7]$$

where $\hat{a}_i^\dagger$ is a creation operator for (spin-)orbital $i$, and $\hat{a}_i$ is the corresponding annihilation operator. Unless otherwise noted, we consider restricted sets of orbitals, so that the one-electron function is independent of other quantum numbers (e.g., spin). For notational ease, the circumflex will be suppressed when there is no chance of confusion.

Although Eq. [2], written in terms of variances, kurtoses and symmetric cokurtoses, is precise, and even elegant, from a statistical perspective, it does not lend itself to computationally straightforward expressions. Explicit expressions for all terms needed to use Eq. [2] for a Riemannian trust-region method using a truncated Conjugate Gradient algorithm for the line search are given in the Appendix. Here we provide derivation of exemplar terms. By direct substitution, Eq. [2] for three-dimensional Euclidean space can be written in terms of cross terms of Cartesian coordinates, e.g.,

$$\langle p|v_x^2 v_y^2|p\rangle = \langle p|x^2y^2|p\rangle - 2\langle p|x^2y|p\rangle\langle p|y|p\rangle - 2\langle p|xy^2|p\rangle\langle p|x|p\rangle$$
$$+ \langle p|x^2|p\rangle\langle p|y|p\rangle^2 + \langle p|x|p\rangle^2\langle p|y^2|p\rangle \quad [8]$$
$$- 3\langle p|x|p\rangle^2\langle p|y|p\rangle^2 + 4\langle p|xy|p\rangle\langle p|x|p\rangle\langle p|y|p\rangle$$

and diagonal terms, e.g.,

$$\langle p|v_x^4|p\rangle = \langle p|x^4|p\rangle - 4\langle p|x^3|p\rangle\langle p|x|p\rangle + 6\langle p|x^2|p\rangle\langle p|x|p\rangle^2 - 3\langle p|x|p\rangle^4 \quad [9]$$

which can be organized into,

$$\mu_4^p$$
$$= \sum_{t=\{1,2,3\}} \left[ \begin{array}{c} \langle p|x_t^4|p\rangle - 4\langle p|x_t^3|p\rangle\langle p|x_t|p\rangle + 6\langle p|x_t^2|p\rangle\langle p|x_t|p\rangle^2 \\ -3\langle p|x_t|p\rangle^4 \end{array} \right]$$
$$+ 2\sum_{t>u} \left[ \begin{array}{c} \langle p|x_t^2 x_u^2|p\rangle - 2(1+\hat{P}_{tu})\langle p|x_t^2 x_u|p\rangle\langle p|x_u|p\rangle \\ +(1+\hat{P}_{tu})\langle p|x_t^2|p\rangle\langle p|x_u|p\rangle^2 \\ -3\langle p|x_u|p\rangle^2\langle p|x_t|p\rangle^2 + 4\langle p|x_t x_u|p\rangle\langle p|x_t|p\rangle\langle p|x_u|p\rangle \end{array} \right] \quad [10]$$

where $\hat{P}_{tu}$ is the permutation operator.

The gradient of the cost function, i.e., Eq. [1], at the origin of the chart

$$\left.\frac{\partial \xi_{FM}^m}{\partial \kappa_{ij}}\right|_{\kappa=0} = m \sum_p (\mu_4^p)^{m-1} \left.\frac{\partial \mu_4^p}{\partial \kappa_{ij}}\right|_{\kappa=0} \quad [11]$$

involves terms that depend on the polynomial degree; for a single product term,

$$\left.\frac{\partial}{\partial \kappa_{ij}}\langle p|x^2y^2|p\rangle\right|_{\kappa=0} = 2\delta_{ip}\langle j|x^2y^2|p\rangle - 2\delta_{jp}\langle i|x^2y^2|p\rangle \quad [12]$$

substituting Eq. [12] into [11],

$$\left.\frac{\partial \xi_{FM}^m(\langle x^2y^2\rangle)}{\partial \kappa_{ij}}\right|_{\kappa=0} = 2m\left\{\langle j|x^2y^2|i\rangle(\mu_4^i)^{m-1} - \langle i|x^2y^2|j\rangle(\mu_4^j)^{m-1}\right\} \quad [13]$$

On the other hand, for a double product, e.g. $\langle x^2y\rangle\langle y\rangle$, the contribution is

$$\left.\frac{\partial \xi_{FM}^m(\langle x^2y\rangle\langle y\rangle)}{\partial \kappa_{ij}}\right|_{\kappa=0} = m\sum_p \left\{\left(\frac{\partial\langle x^2y\rangle}{\partial \kappa_{ij}}\bigg|_{\kappa=0}\right)\langle y\rangle + \langle x^2y\rangle\left(\frac{\partial\langle y\rangle}{\partial \kappa_{ij}}\bigg|_{\kappa=0}\right)\right\}(\mu_4^p)^{m-1} \quad [14]$$

where expectation values are over orbital $p$ unless otherwise noted, i.e. $\langle x^2y\rangle \equiv \langle p|x^2y|p\rangle$, gives

$$\left.\frac{\partial \xi_{FM}^m(\langle x^2 y\rangle\langle y\rangle)}{\partial \kappa_{ij}}\right|_{\kappa=0} = 2m(\mu_4^i)^{m-1}\{\langle j|x^2y|i\rangle\langle i|y|i\rangle + \langle i|x^2y|i\rangle\langle j|y|i\rangle\}$$

$$- 2m(\mu_4^j)^{m-1}\{\langle i|x^2y|j\rangle\langle j|y|j\rangle + \langle j|x^2y|j\rangle\langle i|y|j\rangle\} \quad [15]$$

Lastly, this expression must be multiplied by the coefficient of the original term, i.e. a -2 in Eq. [8] or Eq. [10]. Moment integrals, up to the needed hexadecapole integrals, are calculated analytically using Obara-Saika recursion,[19] as implemented in the local code "undmol".[20]

## 3. Truncated conjugate-gradient Riemannian trust-region algorithm

In a classical application of Newton's method, a solution is sought to a local second-order Taylor series $Q(\vec{\kappa})$ of an objective function. For the problem at hand,

$$Q(\vec{\kappa}) = f(\vec{0}) + \vec{\kappa}^T \text{grad} f(\vec{0}) + \frac{1}{2}\vec{\kappa}^T \text{Hess } f(\vec{0})\vec{\kappa} \quad [14]$$

where $f(\vec{0})$ is the value of the cost function $\xi_{FM}^m(\vec{\kappa})$ at the expansion point $\boldsymbol{\kappa} = \mathbf{0}$, with dependence on unique elements (cf. Eqs. [1] and [10]) emphasized, $\text{grad} f(\vec{0})$ and $\text{Hess } f(\vec{0})$ are the gradient and Hessian of the objective function $\xi_{FM}^m(\vec{0})$, respectively. For notational convenience, let us define $g \equiv \text{grad } f(\vec{0})$ and $H \equiv \text{Hess } f(\vec{0})$; we will also omit the half-headed arrow in $\vec{\kappa}$ whenever there is no ambiguity in meaning.

Newton's method works well provided that (*i*) all directions have positive curvature (i.e., the Hessian is positive definite) and (*ii*) the curvature of different directions does not vary too much. When either (or both) of these conditions are not met, a modification of the Newton method must be made to ensure progress towards a minimum.[21] Modification of the Newton method fall into two broad categories; one in which one or more eigenvalues of the Hessian is shifted (i.e., giving so-called trust-region methods) and another that performs a line search for an update vector along the negative derivative of the objective function.[21] Both approaches are successful, and have been described in the optimization literature, generally by advocates with fervor rarely seen outside of White Sox vs. Cubs fans. The efficiency of either depends on details of the function to be optimized. Høyvik and Jørgensen base their treatment of the fourth order method on a trust-region method and shift the diagonal.[11] This strategy is ideal for situations in which there are few negative eigenvalues and an empty or low-dimensional null space; in pragmatic terms, a good starting guess is available. In contrast, line-search methods are capable of delivering meaningful large steps, regardless of the Sylvester inertia of the Hessian. However, line-search methods are prone to excessive computational effort for low curvature directions, and actual implementations use sophisticated modifications, such as hook steps or double dogleg steps. In consideration of optimization on a Riemannian manifold, which is at best approximately Euclidean over a limited range in which a chart is valid, and which requires potentially computationally expensive pullbacks, Absil et al. explore inexact line searches within a trust region. The defined region of the trust-region method is a hypersphere of radius $\Delta$ around the expansion point, where $\|\vec{\kappa}\| = \sqrt{\vec{\kappa}^T \vec{\kappa}} \leq \Delta$, and management of retractions play a key role.[17] To minimize the fourth moment cost function, $\xi_{FM}^m$ in Eqs. [1] and [10], we used truncated Conjugate-Gradient Riemannian Trust-Region (RTR-tCG) algorithm of Absil *et al.*[17] The RTR-tCG method includes two parts, updating the current iterate and solving a *trust-region subproblem*.

Within the trust-region subproblem, the *truncated Conjugate-Gradient* (tCG) method is used to compute an approximate minimizer κ of the model, which is dubbed the *inner iteration*. As a reminder, the symbol $\kappa$ is used to denote the unique components of the vector, $\vec{\kappa}$, or the skew symmetric vector, $\boldsymbol{\kappa}$, on the tangent manifold, which should be clear from context. When the Hessian is positive definite for iteration $j$, $(\kappa^T H \kappa) > 0$, the tCG algorithm works the same as the traditional CG, however, whenever the Hessian is not positive definite, or if the Newton step exceeds the trust radius, then it looks for a $\kappa$ on the boundary of the trust-region to minimize the value of the second-order Taylor series $Q(\kappa)$ in Eq. [14] using Algorithm 11 of Absil et al.[17] With notation adapted to our problem,

1. *Initialization:* Set $\kappa_0 = 0, r_0 = g, \delta_0 = -r_0; j = 0$;

Note that $\kappa$ acquires a subscript that is irrelevant outside the tCG subproblem. Parenthetically, we emphasize that vector quantities in this subproblem are ordinary $n(n-1)/2$ dimensional vectors, and the Hessian matrix has row and column dimension of $n(n-1)/2$ dimensions.

2. *loop*
3.     *Negative curvature special case:* if $\left(\delta_j^T H \delta_j\right) \leq 0$, then
4:         Compute $\tau$ such that $\eta = \eta_j + \tau \delta_j$ minimizes $Q(\eta)$ in

$$Q(\eta) = f(0) + \eta^T g + \frac{1}{2}\eta^T H\eta \text{ and satisfies } \|\eta\| = \sqrt{\eta^T\eta} = \Delta;$$

5:         return $\eta$;  
6:         end if  
7:         Set $\alpha_j = \dfrac{(r_j^T r_j)}{(\delta_j^T H \delta_j)}$;  
8:         Set $\eta_{j+1} = \eta_j + \alpha_j \delta_j$;  
9:         *Boundary special case:* if $\|\eta_{j+1}\| \geq \Delta$, then  
10:           Compute $\tau \geq 0$ such that $\eta = \eta_j + \tau\delta_j$ satisfies $\|\eta\| = \Delta$;  
11:           return $\eta$  
12:           end if  
13:         *Begin Conjugate Gradient update:* Set $r_{j+1} = r_j + \alpha_j H \delta_j$  
14:         If a stopping condition is satisfied, return $\eta_{j+1}$  
15:         Set $\beta_{j+1} = \dfrac{r_{j+1}^T r_{j+1}}{r_j^T r_j}$  
16:         $\delta_{j+1} = -r_{j+1} + \beta_{j+1}\delta_j$  
17:         Set $j = j + 1$;  
18. *end loop*

The simplest stopping criterion of the loop within the tCG algorithm is to truncate after a predetermined number of iterations. Of course, it is impossible to know this value, unless a specific problem has been done before, and this stopping criterion is primarily to guard against a runaway calculation. Note that the tCG algorithm does not solve simultaneous equations or eigenvectors and thus immune to usual defects in the Hessian. However, this comes at a price: the analysis of convergence of RTR-tCG is nontrivial. Absil et al. (§7.4)[17] analyze convergence with the result that choosing the stopping criterion

$$\|r_{j+1}\| \leq \|r_0\| \min(\|r_0\|^\theta, \gamma) \qquad [15]$$

where $\theta > 0$ and $\gamma$ are real parameters ($\theta = 0.5, \gamma = 0.001$) guarantees superlinear convergence. In fourth and tenth steps of the algorithm, $\tau$ is determined by calculating the positive root of the quadratic equation

$$\tau^2\left(\delta_j^T \delta_j\right) + 2\tau(\eta^T \delta_j) = \Delta^2 - (\eta^T \eta). \qquad [16]$$

Once the tCG iteration returns the new vector $\vec{\kappa}$, the new point on the Riemannian manifold is obtained by generating a vector on the tangent manifold, $\kappa$, and then retracting exp(-$\kappa$). In terms of the basis orbitals,

$$(|\tilde{p}\rangle|\tilde{q}\rangle \ldots |\tilde{s}\rangle) = (|p\rangle|q\rangle \ldots |s\rangle)\exp(-\kappa) \qquad [17]$$

Because we update the chart after each successful displacement on the Riemannian manifold, the molecular orbitals are updated and $\kappa \equiv 0$ (cf. discussion around Eq. [6]). As a result, $\kappa$ is always small, and the usual eigenvector-based exponentiation is accurate, although we have verified the accuracy using the Padé-based method.[22] For surety in our numerical trials, we orthonormalize the new molecular orbitals (using Householder-based QR), but this step is generally unnecessary.

Upon completion of the tCG subproblem, control is returned to the Riemannian Trust Radius main problem. In iteration $k$, a determined step $\kappa_k$ yields the function value $(\xi_m)_{k+1}$ of iteration $k + 1$. This value is predicted by the second-order expansion in previous iteration, $k$, $Q(\kappa)$, cf. Eq. [14]. The quality of the model $Q(\kappa)$, in iteration $k + 1$, is evaluated by the ratio between the observed and predicted change in the target function $\xi_m$,

$$\rho_k = \frac{(\xi_m)_{k+1} - \left(\xi_m^{[0]}\right)_k}{Q(\kappa_k) - \left(\xi_m^{[0]}\right)_k} \qquad [18]$$

The closer that $\rho_k$ is to unity, the more accurate is the quadratic approximation. If $\rho_k > (\rho' = 0.05)$ then the step and the new chart (see above) is accepted, otherwise the step is rejected, and optimization returns to the old chart but with a decreased trust radius.

Whether a step is accepted or not, the trust radius is updated based on how quadratic the surface is,

$$\Delta_{k+1} = \frac{1}{4}\Delta_k \quad if \quad \rho_k \leq 0.25 \qquad [19]$$

$$\Delta_{k+1} = \min(2\Delta_k, \Delta_{\max}) \quad if \quad \rho_k \geq 0.75 \text{ and } \|\kappa\| = \sqrt{\kappa_k^T \kappa_k} = \Delta_k \qquad [20]$$

otherwise $\Delta_{k+1} = \Delta_k$.[17]

## 4. Results

In this section, the results of RTR-tCG optimization of fourth moment cost function implemented in our local quantum chemistry code UNDMOL are presented. Our localization algorithm can rotate any specified subset of orbitals, such as orbitals that are invariant to rotations in the underlying Hamiltonian. Numerical tests are shown for RHF calculations on $H_{10}$, in various geometries, and on cyclobutadiene. A second set of tests are shown for the active space in MCSCF calculations on cyclobutadiene and on the ground and lowest excited states of propargyl. There are two situations illustrating intrinsic difficulties of localization in the virtual orbital space. The first scenario occurs when the atomic orbital (AO) basis is large and thus more linearly dependent than usual, which makes orthonormal localized molecular orbitals hard to achieve. On the other hand, when the AO basis has essentially no overlap, the isosurfaces of any localizing function may become more jagged.[6]

## 4.1 Application of localization on the occupied and the virtual blocks of $H_{10}$ in various geometries

As a first benchmark for assessing the efficacy of the RTR-tCG algorithm, a set of $H_{10}$ in one, two, and three dimesions[23] were examined (cf. Figure 1), because these structures allow an examination of sensitivity of FM localization to number of nearest neighbors. Note that the localization methods remove the symmetry of canonical molecular orbitals, and we did not use symmetry during any of the computations. Considering the 1.50 Å H-H distance for all models, we employed different basis sets [the minimal basis set STO-3G, 6-31G, and 6-31$^{++}$G$^{**}$], to address this question; does this algorithm retain similar quality for the growing size of the basis sets in different dimensional models?

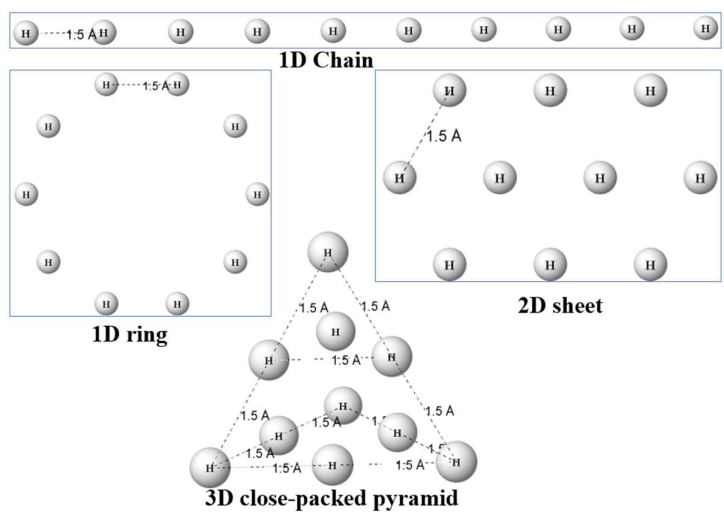

Figure 1. Structure of $H_{10}$ model systems examined in this study. The nearest-neighbor H-H distance is 1.5 Å.

### 4.1.1. Quantitative and Qualitative analysis of $H_{10}$ orbitals

The second moment minimization (or the Boys localization) directly confines the bulk of the orbitals to a small volume in space, in contrast, the fourth central moment approach targets on the tail region of the orbitals.[3] Hence, The fourth root of the fourth central moment (fourth moment orbital spread)[11] is used to represent the locality of an orbital,

$$\sigma_4^p = (\mu_4^p)^{1/4} \qquad [21]$$

One measure of how well a set of orbitals as a whole is local or not, the average locality, defined as

$$\sigma_4^{ave} = \frac{1}{N}\sum_{p=1}^{N} \sigma_4^p \qquad [22]$$

where $N$ is the number of orbitals in the set, could be used. However, reporting the locality of the least local orbital (maximum orbital spread, $\sigma_4^{max}$) often characterizes the locality much better.[3] Figure 2 through Figure 5 present the spreads of LMOs relative to the CMOs, and indicates qualitatively and quantitatively that FM localization employed by RTR-tCG optimization performs quite well for the minimal basis set (STO-3G), regardless of geometry. As larger basis sets are used (cf. Table ), including diffuse functions (6-31$^{++}$G$^{**}$), although the virtual space becomes more complicated (especially with large dimensionality), the algorithm behaves consistently. Figure 2, in particular, shows that the algorithm is able to minimize the maximum orbitals spreads of 1D chain, where $\sigma_4^{max} = 10.18$ decreased to $4.40\ au$ within the occupied orbitals and $\sigma_4^{max} = 10.45$ to $4.42\ au$ for the virtual ones. These localities occurred for the 3D pyramid (Figure 5); for the occupied block from $\sigma_4^{max} = 3.71$ to $3.00\ au$, and the virtual block from $\sigma_4^{max} = 3.54$ to $3.02\ au$. Table information paints a similar picture: the FM localization with RTR-tCG algorithm is able to properly deal with the more challenging virtual spaces. For the 1D chain, the virtual orbital spread for CMOs increases from $\sigma_4^{max} = 10.5\ au$ for the minimal basis set (STO-3G) to $15.9\ au$ for the 6-31$^{++}$G$^{**}$ one, while the algorithm confines the spreads to $\sigma_4^{max} = 4.3$ and $5.3\ au$, respectively. The virtual space of 3D pyramid follows the same pattern of behavior; from the STO-3G to the 6-31$^{++}$G$^{**}$, the orbital extension increases from $\sigma_4^{max} = 3.5$ to $8.6\ au$, while the RTR-tCG FM localization method reduces them to $\sigma_4^{max} = 3.0$ and $4.6\ au$, respectively.

| CMOs | $\sigma_4^p$ | LMOs | $\sigma_4^p$ |
|---|---|---|---|
| | 7.17 | | 4.31 |
| | 9.30 | | 3.97 |
| | 10.18 | | 3.97 |
| | 10.41 | | 4.32 |
| | 9.47 | | 4.40 |
| | 9.49 | | 4.42 |
| | 10.45 | | 4.32 |
| | 10.28 | | 3.99 |
| | 9.48 | | 3.99 |
| | 7.38 | | 4.32 |

Figure 2. The isosurfaces of CMOs and LMOs for chain $H_{10}$ ($m = 2$) were plotted using the iso value of 0.06 (obtained at a restricted Hartree-Fock (RHF) level by using STO-3G basis set). Their average fourth moment orbital spreads ($au$) are also presented.

| CMOs | | LMOs | |
|---|---|---|---|
| 4.90 | 4.91 | 3.29 | 3.28 |
| 4.91 | 4.94 | 4.68 | 4.68 |
| 4.94 | 4.95 | 3.84 | 4.55 |
| 4.95 | 4.94 | 4.94 | 3.38 |
| 4.94 | 4.93 | 4.55 | 3.38 |

Figure 3. The isosurfaces of CMOs and LMOs for ring $H_{10}$ ($m = 2$) obtained at RHF/STO-3G level.

| CMOs | | LMOs | |
|---|---|---|---|
| 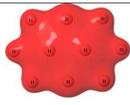 3.29 | 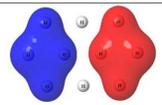 4.09 | 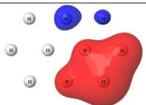 3.13 | 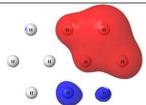 3.13 |
| 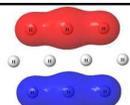 3.67 | 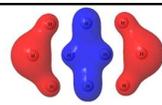 4.32 | 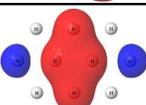 3.35 | 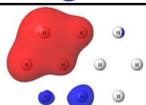 3.13 |
| 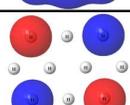 4.18 | 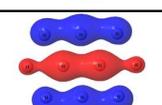 3.47 | 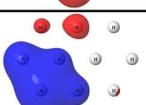 3.13 | 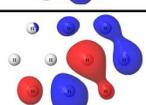 3.26 |
| 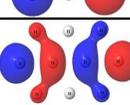 4.23 | 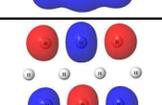 3.82 | 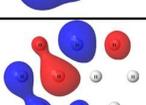 3.26 | 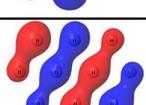 3.68 |
| 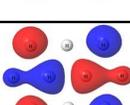 3.75 | 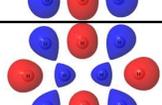 4.06 | 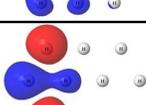 3.03 | 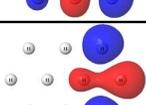 3.03 |

Figure 4. The isosurfaces of CMOs and LMOs for sheet $H_{10}$ ($m = 2$) obtained at RHF/STO-3G level.

| CMOs | | LMOs | |
|---|---|---|---|
| 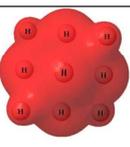 2.89 | 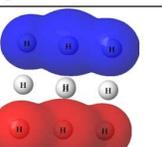 3.41 | 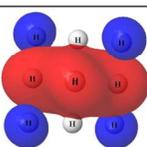 3.00 | 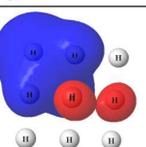 2.73 |
| 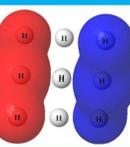 3.43 | 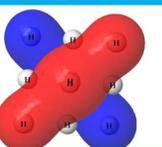 3.43 | 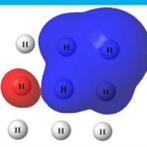 2.73 | 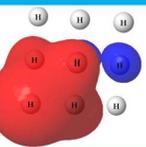 2.73 |
| 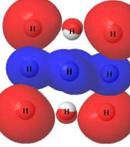 3.71 | 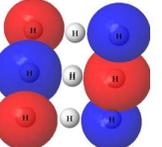 3.53 | 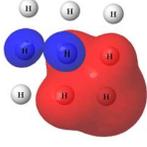 2.73 | 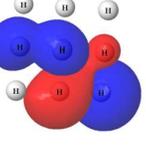 3.01 |
| 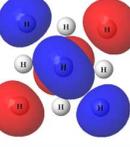 3.53 | 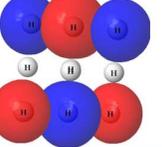 3.54 | 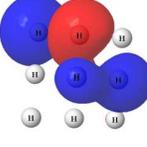 3.00 | 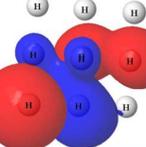 3.02 |
| 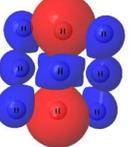 3.04 | 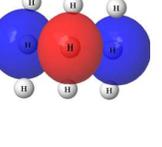 2.73 | 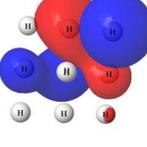 3.00 | 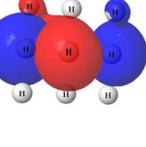 2.78 |

Figure 5. The isosurfaces of CMOs and LMOs for pyramid $H_{10}$ ($m = 2$) obtained at RHF/STO-3G level.

Table 1. Maximum orbital spreads, $\sigma_4^{max}(au)$, of CMOs and LMOs for H$_{10}$ set at HF level with different basis sets (1) STO-3G, (2) 6-31G, and (3) 6-31$^{++}$G$^{**}$ ($m = 2$).

|  |  | Chain | | | Ring | | | Sheet | | | Pyramid | | |
|---|---|---|---|---|---|---|---|---|---|---|---|---|---|
|  |  | (1) | (2) | (3) | (1) | (2) | (3) | (1) | (2) | (3) | (1) | (2) | (3) |
| Occ | CMO | 10.4 | 10.5 | 10.5 | 4.9 | 5.1 | 5.2 | 4.3 | 4.6 | 4.6 | 3.7 | 4.3 | 5.1 |
| Occ | LMO | 4.3 | 3.9 | 4.0 | 3.9 | 5.1 | 4.1 | 3.1 | 3.5 | 3.5 | 2.7 | 3.1 | 3.5 |
| Vir | CMO | 10.5 | 13.1 | 15.9 | 5.0 | 6.0 | 10.2 | 4.2 | 5.4 | 8.7 | 3.5 | 4.7 | 8.6 |
| Vir | LMO | 4.3 | 3.8 | 5.3 | 4.6 | 3.6 | 5.4 | 3.3 | 3.3 | 4.3 | 3.0 | 3.0 | 4.6 |

### 4.1.2. Impact of different penalty values on H$_{10}$ set

Increased powers of exponents were introduced by Jansik et al.[4] to increase penalties for delocalization. We examine the impact of increasing the penalty (value of $m$) on the orbitals on the RTR-tCG algorithm and compare with CMOs. Generally speaking, Table signifies that the orbital spread distribution for different dimensions of H$_{10}$ set gets narrower when $m$ changes from one to two, however, increasing the penalty exponent to three and four leads to the larger orbital spreads. This somewhat puzzling result can be understood by recalling that the Fourth Moment function minimizes the sum of orbital spreads, and not the maximum orbital spread that is tabulated in Table . Therefore, these results corroborate the observation of Jansik et al., that the value of 2 for the penalty exponent provides improvement in localization, but there is no reason to use higher exponents.[4] It should also be appreciated that larger exponents will exacerbate numerical errors, although this wasn't an issue for the test calculations since long double precision was used when there were questions of numerical errors.

Table 2. Maximum orbital spreads, $\sigma_4^{max}(au)$, of CMOs and LMOs for the cc-pVDZ basis set[24] using different penalty values

| $m$ | Chain | | Ring | | Sheet | | Pyramid | |
|---|---|---|---|---|---|---|---|---|
|  | Occ | Vir | Occ | Vir | Occ | Vir | Occ | Vir |
| 1 | 3.67 | 3.38 | 4.67 | 3.53 | 3.02 | 3.43 | 3.03 | 3.30 |
| 2 | 3.95 | 3.43 | 4.67 | 3.49 | 3.18 | 2.77 | 3.16 | 2.75 |
| 3 | 4.03 | 3.29 | 4.67 | 3.64 | 3.25 | 2.89 | 3.18 | 2.67 |
| 4 | 4.07 | 3.40 | 4.67 | 3.98 | 3.39 | 2.80 | 3.19 | 2.75 |
| CMO | 10.48 | 12.34 | 5.18 | 6.45 | 4.59 | 5.61 | 4.56 | 5.00 |

## 4.2 Cyclobutadiene (c-C$_4$H$_4$)

Cyclobutadiene was the second case study to test the approach. Due to the presence of two identical configurations of C-C double bonds, the transition state of the cyclobutadiene automerization displays a prominent multireference character leading to a demanding description of its electronic structure methods. Lyakh et al. applied the coupled-cluster singles, doubles, and perturbative triples [CCSD(T)] to calculate the potential energy surface of the cyclobutadiene automerization.[25] The C-H bond lengths 1.079 Å, and H-C-C angles 135° were kept the same for the minimum and transition state of c-C$_4$H$_4$, and they relaxed the restriction on C-C bond lengths to optimize their geometries at CCSD(T) level of theory using the spherical cc-pVTZ basis set.[24] We used the optimized geometries (Figure 2) for our calculations.

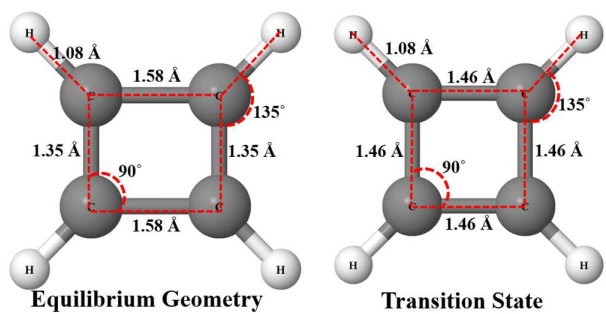

Figure 2. Geometries of the rectangular ground-state (D$_{2h}$) and square transition (D$_{4h}$) structure of cyclobutadiene (c-C$_4$H$_4$) optimized at the CCSD(T)/cc-pVTZ level[25]

### 4.2.1. Application of localization on the occupied and the virtual blocks of c-C$_4$H$_4$

In this scheme, the orbital space, generated at HF/cc-pVTZ, is divided into the occupied and virtual orbitals, and the RTR-tCG technique was applied separately on each orbital block. Localized MOs reproduce canonical HF energies with the error less than $10^{-6} mE_h$, and in Figure 1 and **Error! Reference source not found.**, the $\sigma$ and $\pi$ attributes in LMOs are preserved to a large extent. For instance, the c-C$_4$H$_4$ minimum (Figure 1) shows that the produced LMOs feature $\sigma$ C-H bonds, although there is a mixing of $\sigma$ and $\pi$ C-C bonds.

| Occupied | | Equilibrium Geometry | | | | |
|---|---|---|---|---|---|---|
| CMOs | | 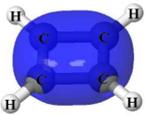 | 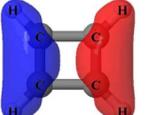 | 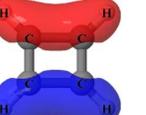 | 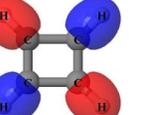 | 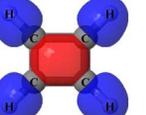 |
| | $\sigma_4^p$ | 2.39 | 3.14 | 3.39 | 3.83 | 3.72 |
| | | 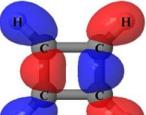 | 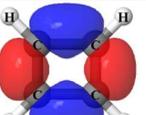 | 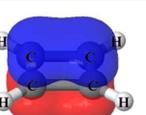 | 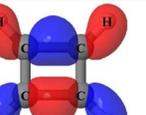 | 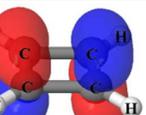 |
| | $\sigma_4^p$ | 3.61 | 2.84 | 2.93 | 3.53 | 3.26 |
| LMOs | | 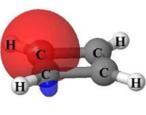 | 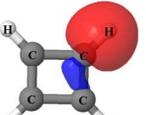 | 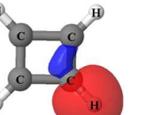 | 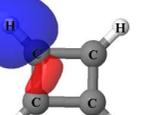 | 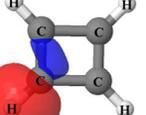 |
| | $\sigma_4^p$ | 2.07 | 2.08 | 2.09 | 2.08 | 2.28 |
| | | 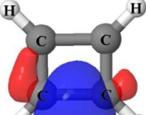 | 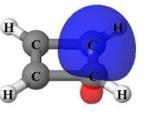 | 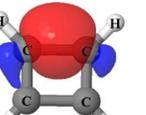 | 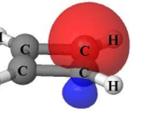 | 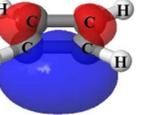 |
| | $\sigma_4^p$ | 2.48 | 2.22 | 2.14 | 2.26 | 2.39 |

| Virtual | $\sigma_4^{max}$ |
|---|---|
| CMOs | 6.99 |
| LMOs | 3.77 |

Figure 1. Using the iso value of 0.06, the isosurfaces of CMOs and LMOs of $c$-C$_4$H$_4$ minimum ($m = 2$) are presented (obtained at HF/cc-pVTZ level). The fourth moment orbital spreads ($au$) are also provided.

| Occupied | Transition State | | | | |
|---|---|---|---|---|---|
| | 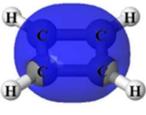 | 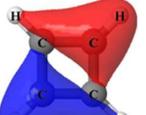 | 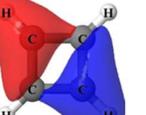 | 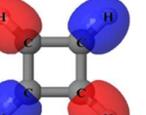 | 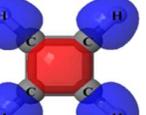 |
| CMOs $\sigma_4^p$ | 2.33 | 3.27 | 3.22 | 3.82 | 3.72 |
| | 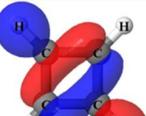 | 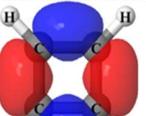 | 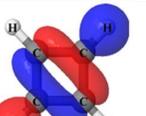 | 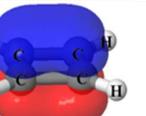 | 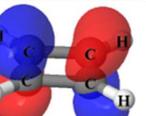 |
| $\sigma_4^p$ | 3.53 | 2.83 | 3.56 | 2.93 | 3.36 |
| | 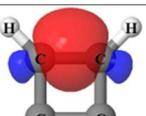 | 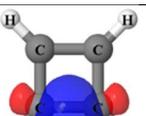 | 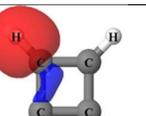 | 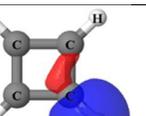 | 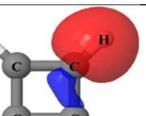 |
| LMOs $\sigma_4^p$ | 2.14 | 2.07 | 2.44 | 2.08 | 2.08 |
| | 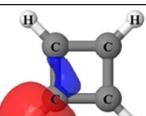 | 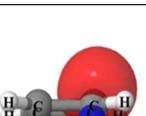 | 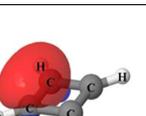 | 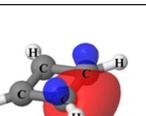 | 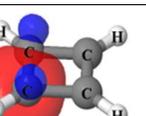 |
| $\sigma_4^p$ | 2.14 | 2.44 | 2.07 | 2.44 | 2.44 |

| Virtual | $\sigma_4^{max}$ |
|---|---|
| CMOs | 6.99 |
| LMOs | 3.71 |

Figure 8. Using the iso value of 0.06, the isosurfaces of CMOs and LMOs of $c$-C$_4$H$_4$ transition state ($m = 2$) are presented (obtained at HF/cc-pVTZ level). The fourth moment orbital spreads ($au$) are also provided.

### 4.2.2. Application of localization on an active space of $c$-C$_4$H$_4$

In the following study of $c$-C$_4$H$_4$, the optimization of molecular orbitals in the multiconfigurational self-consistent-field (MCSCF) with cc-pVTZ basis set was carried out on these sets of orbitals: the four lowest molecular orbitals (1a$_g$, 1b$_{2u}$, 1b$_{1u}$, 1b$_{3g}$), which are the 1s orbitals of four carbon atoms, the valence orbitals (2a$_g$, 2b$_{2u}$, 2b$_{1u}$, 4a$_g$, 2b$_{3g}$, 3a$_g$, 3b$_{2u}$, 3b$_{1u}$), which are related to $\sigma_{CC}$ and $\sigma_{CH}$, are taken as the frozen core (i.e., in all configurations of the MCSCF, they are doubly occupied); and (1b$_{3u}$, 1b$_{1g}$, 1b$_{2g}$, 1a$_u$), which are associated with $\pi_{CC}\pi_{CC}^*$, are considered as active space, four electrons in four orbitals (4e, 4o) (cf. Figure 9). Keeping four electrons in $\pi_{CC}\pi_{CC}^*$ for MCSCF calculations, the configurations in this model space generate 8 configuration state functions (CSFs) of $^1A_g$ symmetry and singlet spin symmetry. With $D_{2h}$ symmetry for ground-state and transition-state of $c$-$C_4H_4$, the molecular orbitals were optimized for the lowest singlet $^1A_g$ state.

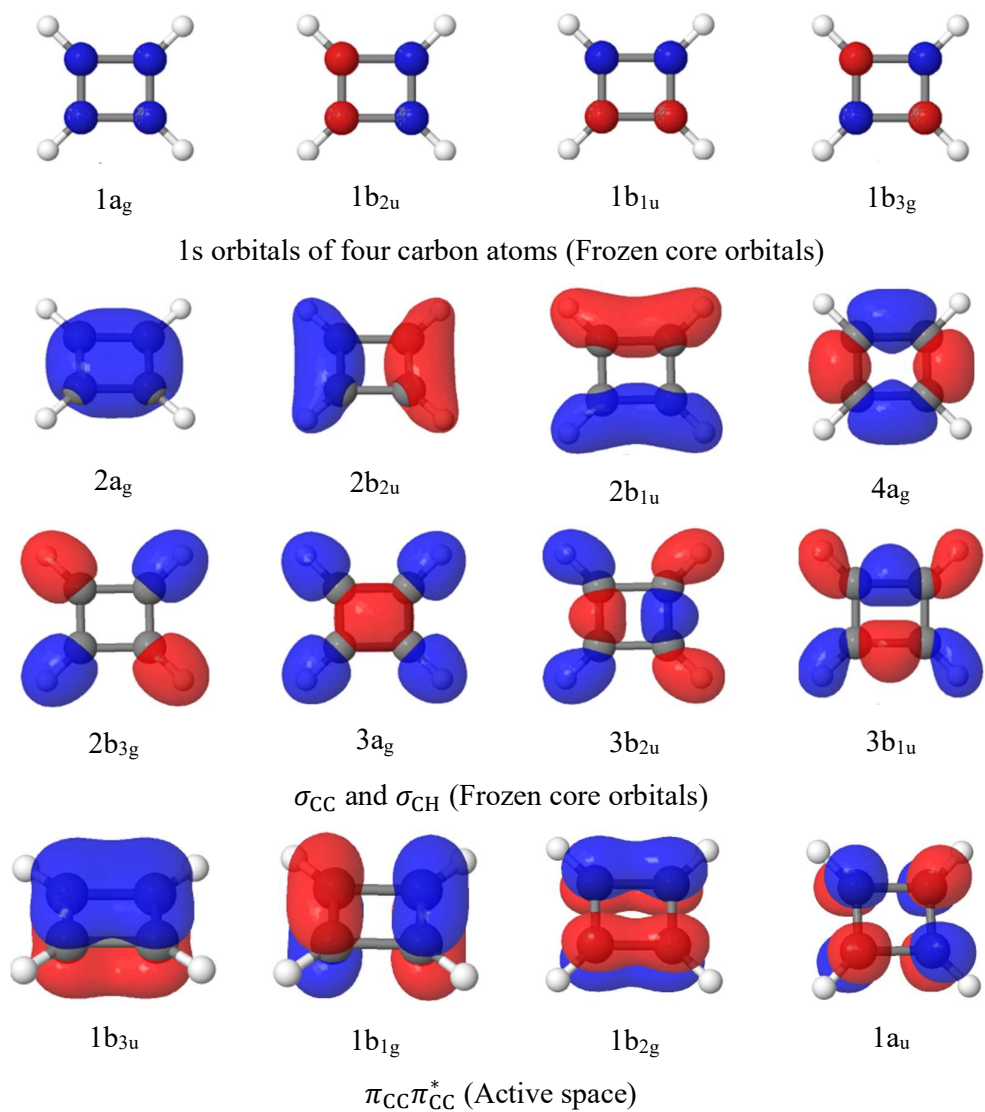

Figure 3. Frozen core and active orbitals of the rectangular ground-state $c$-$C_4H_4$ in MCSCF calculations

Having calculated CMOs at MCSCF/cc-pVTZ level, the localization algorithm was only performed on the active space (4e,4o). After obtaining LMOs of the active orbitals, we ran an MCCI on the final orbitals, in localized form, that is, a single cycle of MCSCF on a converged set of localized orbitals was applied with disabling diagonalization. These calculations on LMOs of the ground-state and transition structure of $c$-$C_4H_4$ show a substantial multiconfigurational nature for their ground state wave functions (Table ), along with 20 CSFs of this model space of $^1A$ symmetry. In each case, the weight of the leading configuration does not exceed 0.72, therefore the cumulative weight of several configurations for each case proves to be crucial. Those localized configurations with the maximum weight for the minimum (0.72) and transition state (0.68) are presented in Table . It is interesting to note that the localization approach produces $p_z$-like orbital for each carbon atom (Figure 4 and **Error! Reference source not found.**), and we did anticipate these results for the transition state but not for the equilibrium geometry of $c$-C$_4$H$_4$. In Figure 4 and Figure 5, values specified for the canonical MO are eigenvalues of the active space subblock of the average Fock matrix, and the MO are corresponding eigenvectors. For the localized MO, the values are the diagonal elements of the average Fock matrix in the localized orbital basis.

Table 3. Leading electron configurations for the $c$-C$_4$H$_4$ at the MCSCF/cc-pVTZ level (4e,4o) for $^1$A symmetry

|  | amplitudes | configuration |
|---|---|---|
| Equilibrium state | +0.72 | $(13a)^1_\uparrow (14a)^1_\downarrow (15a)^1_\uparrow (16a)^1_\downarrow$ |
|  | -0.31 | $(13a)^0 (14a)^2_{\uparrow\downarrow} (15a)^1_\uparrow (16a)^1_\downarrow$ |
|  | -0.31 | $(13a)^2_{\uparrow\downarrow} (14a)^0 (15a)^1_\uparrow (16a)^1_\downarrow$ |
|  | -0.31 | $(13a)^1_\uparrow (14a)^1_\downarrow (15a)^0 (16a)^2_{\uparrow\downarrow}$ |
|  | -0.31 | $(13a)^1_\uparrow (14a)^1_\downarrow (15a)^2_{\uparrow\downarrow} (16a)^0$ |
| Transition state | +0.68 | $(13a)^1_\uparrow (14a)^1_\downarrow (15a)^1_\uparrow (16a)^1_\downarrow$ |
|  | +0.39 | $(13a)^1_\uparrow (14a)^1_\uparrow (15a)^1_\downarrow (16a)^1_\downarrow$ |

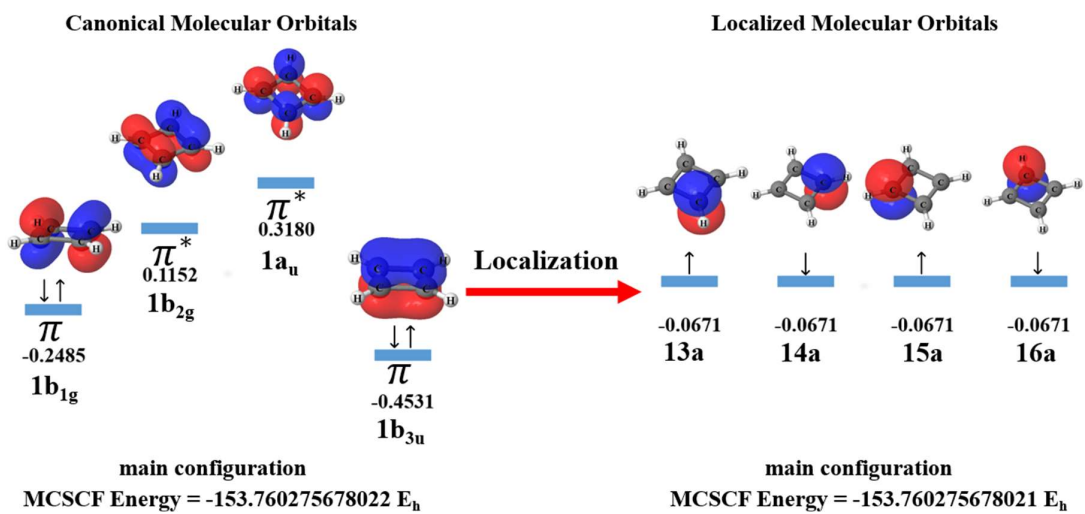

Figure 4. Localization effect on active space of the equilibrium geometry (contour = 0.06)

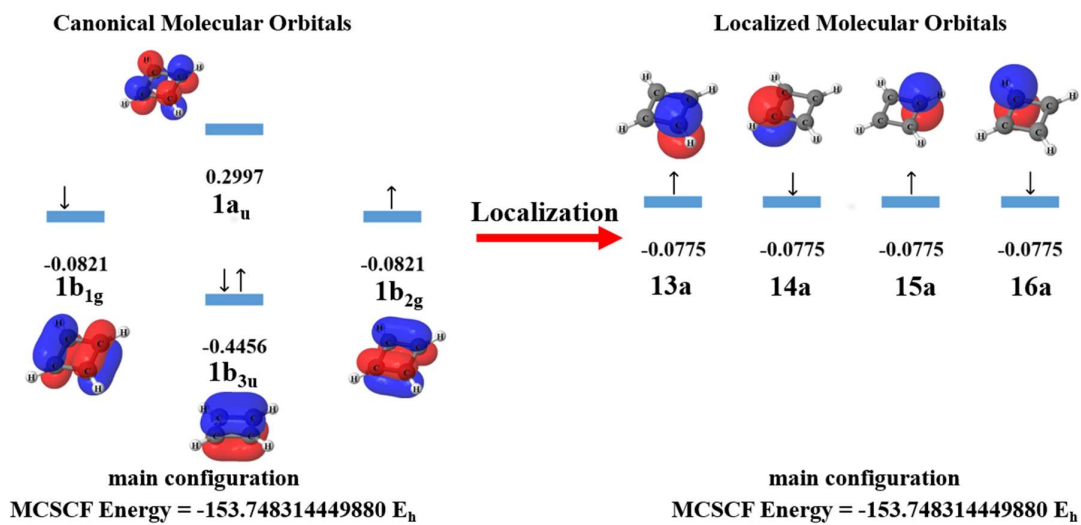

Figure 5. Localization effect on active space of the transition state (contour = 0.06)

### 4.3 The propargyl radical (H$_2$Ċ−C≡CH)

As a final case study, the propargyl radical (H$_2$Ċ−C≡CH), which is the simplest π-conjugated hydrocarbon radical[26] and the most stable among the C$_3$H$_3$ radical structures,[27] was under investigation by the algorithm. Studying the propargyl radical allowed us to examine the performance of the localization algorithm on electronic excited states. The *ab initio* study by Botschwina et al. shows that the ground state of the propargyl radical has C$_{2v}$ symmetry,[28] with a near-prolate asymmetric-top structure. for this study, we use the optimized geometrical parameters from an earlier study by two of the authors at the multireference perturbation method GVVPT2[20,29,31] level using the aug-cc-pVTZ[31] basis set (cf. Figure 6).

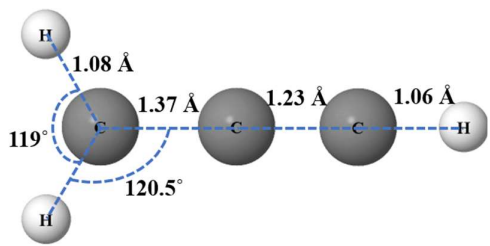

Figure 6. Optimized geometry of the propargyl radical at GVVPT2/aug-cc-pVTZ level.[32]

### 4.3.1. Application of localization on an active space

Before applying the localization, within $C_{2v}$ symmetry, the orbitals $1a_1$-$7a_1$ and $1b_2$ were kept doubly occupied but optimized in the initial MCSCF calculations (cf., Figure 7). In agreement with previous studies, our MCSCF calculations show that the ground electronic configuration is $\ldots 7a_1^2 1b_2^2 1b_1^2 2b_2^2 2b_1^1$, coupled to the many electron $1^2B_1$ state,[26,27,32,33] also that the first and second excited electronic states are $1^2B_2$ and $2^2B_1$, respectively. Table exhibits the canonical active space included five electrons distributed on six orbitals ($1b_1$-$4b_1$ and $2b_2 3b_2$), MCSCF(5e,6o) for $1^2B_1$ (ground electronic state) and $2^2B_1$ (second excited state), and five electrons in five orbitals ($1b_1$-$3b_1$ and $2b_2 3b_2$), MCSCF(5e,5o) for $1^2B_2$ (first excited state).[32]

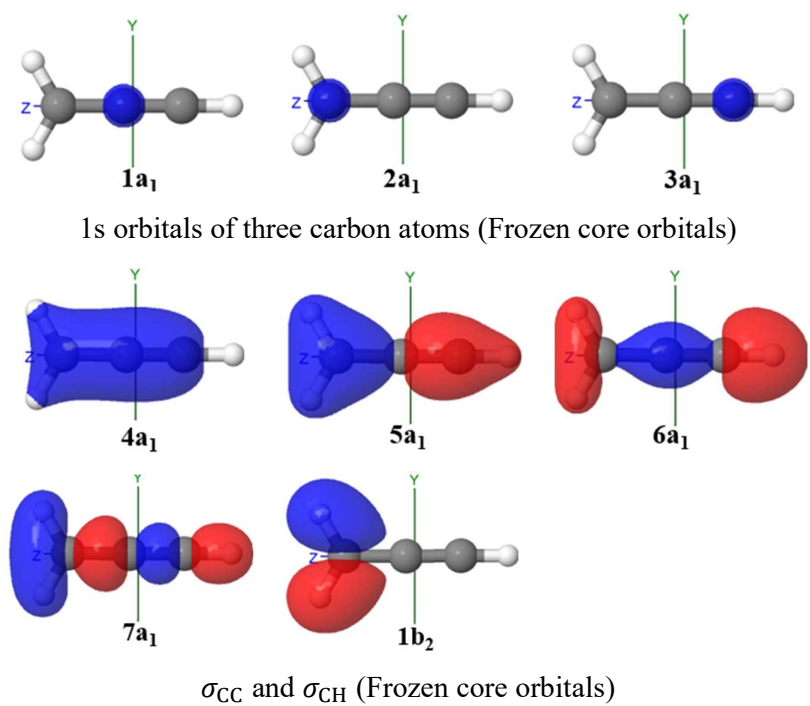

1s orbitals of three carbon atoms (Frozen core orbitals)

$\sigma_{CC}$ and $\sigma_{CH}$ (Frozen core orbitals)

Figure 7. The frozen core orbitals of the propargyl radical

Table 4. Active spaces and electron occupations applied for the propargyl radical at MCSCF/aug-cc-pVTZ level

|  | Ground State | First Excited State | Second Excited State |
|---|---|---|---|
| CMOs | $1^2B_1$ | $1^2B_2$ | $2^2B_1$ |
|  | $(1b_1 2b_1 3b_1 4b_1 2b_2 3b_2)^5$ | $(1b_1 2b_1 3b_1 2b_2 3b_2)^5$ | $(1b_1 2b_1 3b_1 4b_1 2b_2 3b_2)^5$ |
| LMOs | $1^2A$ | $2^2A$ | $3^2A$ |
|  | $(9a 10a 11a 12a 13a 14a)^5$ | $(9a 10a 11a 12a 13a)^5$ | $(9a 10a 11a 12a 13a 14a)^5$ |

In the canonical $1^2B_1$ state, the high occupancy $1b_1$ and $2b_2$ orbitals, along with essentially singly occupied $2b_1$ are $\pi$-type orbitals located predominantly in the CCC backbone of the propargyl radical. The $1b_1$ molecular orbital is formed by the three $p_x$-orbitals associated with three C atoms, the $2b_2$ molecular orbital is made by the two $p_y$-orbitals related to the middle C and terminal C (CH) atoms, and the $2b_1$ molecular orbital is generated by the difference linear combination of the two $p_x$-orbitals clearly related to the terminal C atoms, thereby gaining weakly anti-bonding characteristics (a non-bonding delocalized $\pi$ orbital)[26,35] (cf. Figure ).

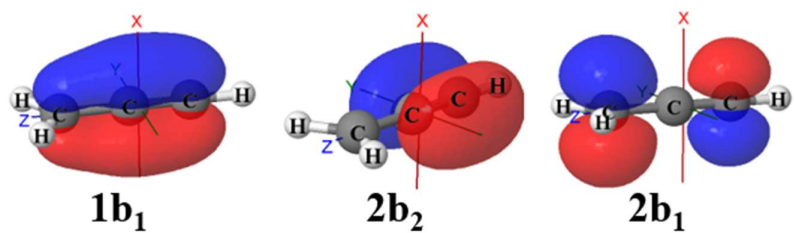

Figure 14. High occupancy canonical MOs in the active space of $1^2B_1$ (ground electronic state)

After completing conventional MCSCF calculations, the localization algorithm was performed on the canonical active spaces to generate localized active orbitals for the ground, first, and second excited states (Table ). These calculations also corroborated that the localization algorithm was implemented correctly, in that the variation in energies between canonical and localized results was less than $10^{-6} mE_h$. Not surprisingly, the wave function is highly multiconfigurational in LMO basis (Table ). In each state, the weight of the leading configuration does not exceed 0.36. For the ground and two excited electronic states, the canonical and localized MOs of their active spaces are pictured in Figure 8 through Figure 10. The natural orbital occupation numbers presented above each LMO, and the diagonal elements of the averaged Fock matrix are given under each MO. It is important to emphasize that the final MCSCF calculations (i.e., actually MCCI) were executed over the LMOs with the semi-canonicalization disabled, because orbital semi-canonicalization would retrieve CMOs from LMOs. It was fascinating to observe that the localization converted the CMOs into LMOs that were helical. In the next section, the nature of the helical orbitals is considered further.

Table 5. Leading localized electron configurations for the propargyl radical at the MCSCF/aug-cc-pVTZ level for $^2A$ symmetry

|  | amplitudes | configuration |
|---|---|---|
| Ground-state (5e,6o) | +0.36 | $(9a)^1_\uparrow(10a)^1_\uparrow(11a)^1_\downarrow(12a)^1_\uparrow(13a)^1_\uparrow(14a)^0$ |
|  | +0.23 | $(9a)^1_\uparrow(10a)^0(11a)^1_\downarrow(12a)^1_\uparrow(13a)^1_\downarrow(14a)^1_\uparrow$ |
|  | −0.22 | $(9a)^1_\uparrow(10a)^1_\uparrow(11a)^1_\downarrow(12a)^1_\uparrow(13a)^0(14a)^1_\uparrow$ |
|  | +0.22 | $(9a)^1_\uparrow(10a)^1_\uparrow(11a)^1_\downarrow(12a)^0(13a)^1_\downarrow(14a)^1_\uparrow$ |
| First excited state (5e,5o) | +0.33 | $(9a)^2_{\uparrow\downarrow}(10a)^1_\uparrow(11a)^1_\downarrow(12a)^1_\uparrow(13a)^0$ |
|  | +0.31 | $(9a)^2_{\uparrow\downarrow}(10a)^1_\uparrow(11a)^1_\uparrow(12a)^0(13a)^1_\downarrow$ |
|  | +0.24 | $(9a)^1_\uparrow(10a)^1_\uparrow(11a)^1_\uparrow(12a)^1_\downarrow(13a)^1_\downarrow$ |
|  | +0.23 | $(9a)^1_\uparrow(10a)^1_\downarrow(11a)^1_\uparrow(12a)^1_\uparrow(13a)^1_\downarrow$ |
|  | +0.21 | $(9a)^2_{\uparrow\downarrow}(10a)^2_{\uparrow\downarrow}(11a)^1_\uparrow(12a)^0(13a)^0$ |
|  | −0.21 | $(9a)^2_{\uparrow\downarrow}(10a)^2_{\uparrow\downarrow}(11a)^0(12a)^1_\uparrow(13a)^0$ |
|  | +0.21 | $(9a)^2_{\uparrow\downarrow}(10a)^1_\uparrow(11a)^2_{\uparrow\downarrow}(12a)^0(13a)^0$ |
|  | +0.21 | $(9a)^2_{\uparrow\downarrow}(10a)^0(11a)^2_{\uparrow\downarrow}(12a)^0(13a)^1_\uparrow$ |
| Second excited state (5e,6o) | +0.24 | $(9a)^1_\uparrow(10a)^1_\uparrow(11a)^1_\downarrow(12a)^1_\uparrow(13a)^1_\downarrow(14a)^0$ |
|  | −0.23 | $(9a)^1_\uparrow(10a)^1_\uparrow(11a)^1_\downarrow(12a)^0(13a)^1_\uparrow(14a)^1_\downarrow$ |

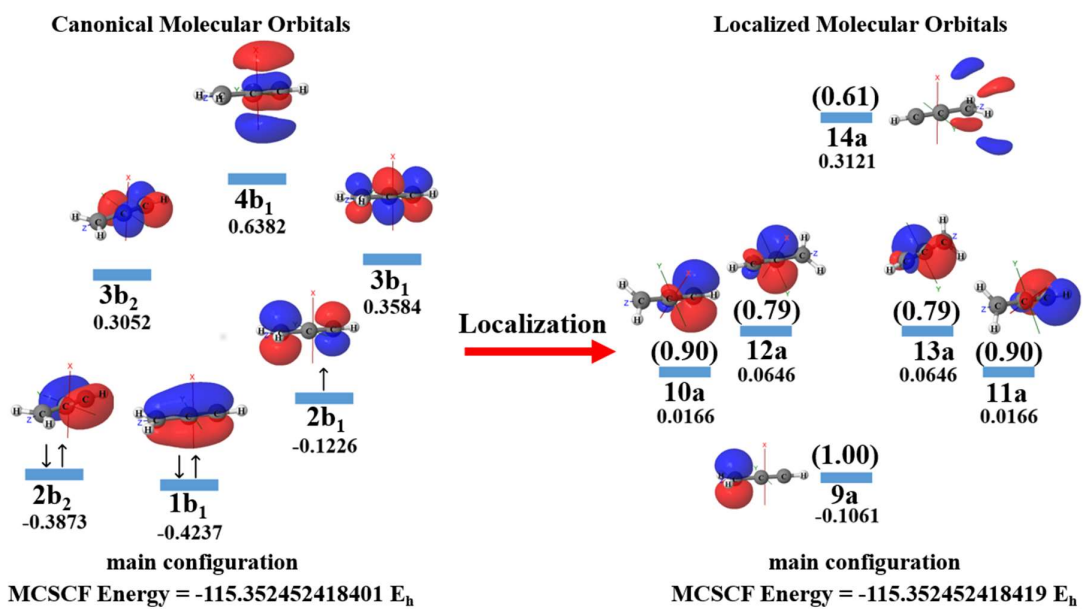

Figure 8. Localization effect on active space of the ground electronic state (contour = 0.06)

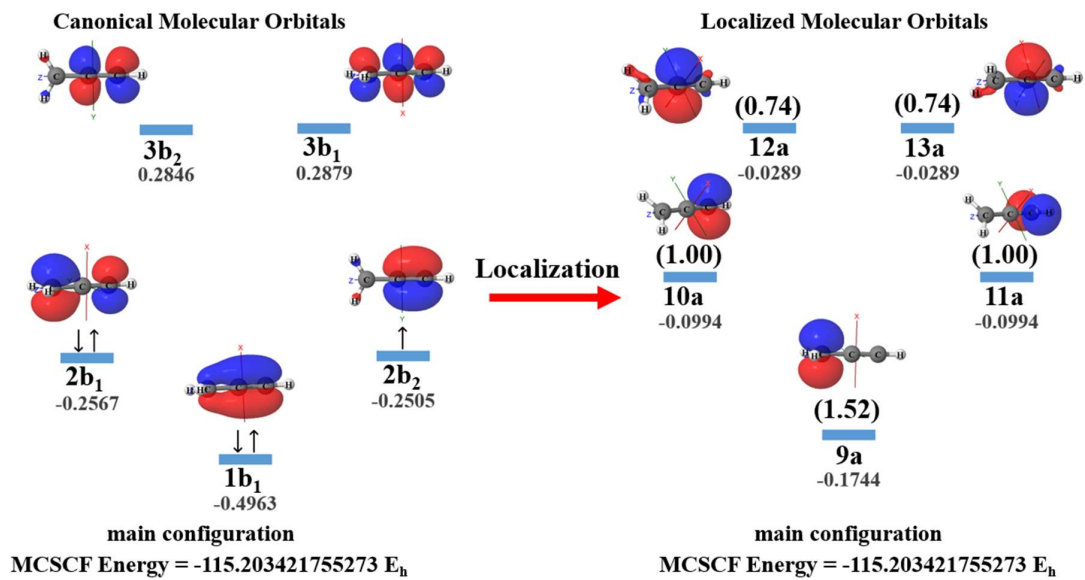

Figure 9. Localization effect on active space of the first excited state (contour = 0.06)

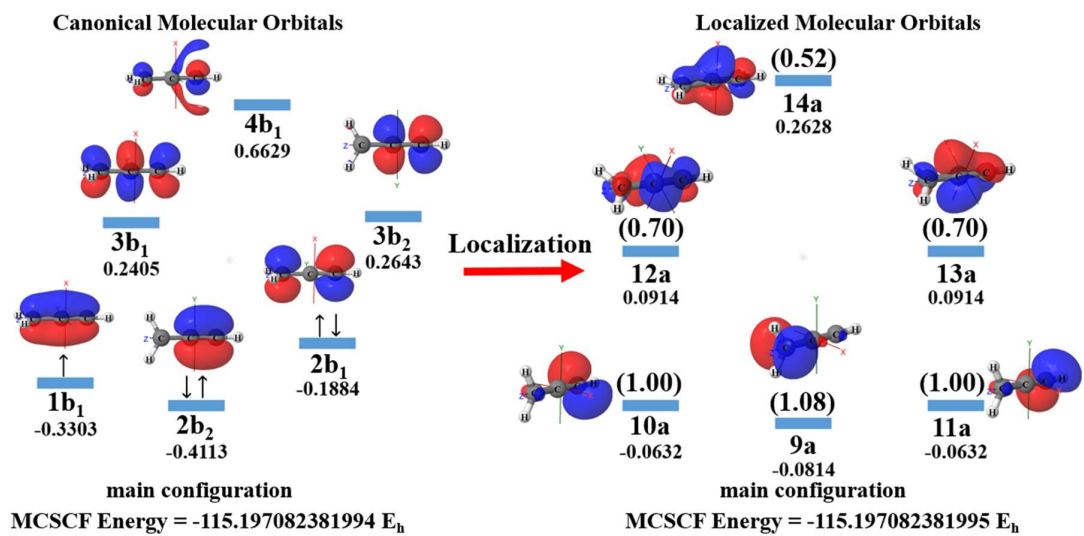

Figure 10. Localization effect on active space of the second excited state (contour = 0.06)

### 4.3.2. Helical molecular orbitals of the propargyl radical

A group of linearly conjugated $\pi$-systems with terminal tricoordinate carbon atoms is dubbed [$n$]cumulenes (Figure 11), where $n$ is the number of cumulated double bonds and the number of carbon atoms is then $n + 1$.[35] For even $n$, closed-shell $\pi$- systems are generated by the mutually perpendicular terminal carbon atoms. Allene is the shortest even [$n$]cumulene: linearly bound three carbons with two double bonds.[36] Because of $D_{2d}$ symmetry of unsubstituted even [n]cumulenes, the frontier $\pi$ molecular orbitals are degenerate, and this symmetry is reduced to $C_2$ or lower by $\alpha,\omega$-disubstitution, so that the orbital degeneracy is consequently lifted.[35]

[n]cumulene, $D_{2d}$

S-$\alpha,\omega$-dimethyl-[n]cumulene, $C_2$

Figure 11. [n]cumulenes[35,36]

For the even [n]cumulenes, where there are odd-carbon atoms (i.e. the allene family),[36] the $\pi$ electrons can be represented by either perpendicular (rectilinear) molecular orbitals or by helical orbitals. The symmetry reduction from $D_{2d}$ to $C_2$, for the $\alpha,\omega$-disubstituted even [n]cumulenes, leads to an obligatory helical representation of the $\pi$ electrons.[35] By applying RTR-tCG algorithm on the propargyl radical, which is a [2]cumulene (three carbon atoms), the symmetry of the orbitals is reduced by the localization from $C_{2v}$ to $C_1$, i.e. a twist of the $\pi$ molecular orbitals appears. Similar to Möbius aromatics, cyclic conjugated systems can be represented by twisted $\pi$ interactions for allenes (Figure 12).[36] These right- and left-handed helices lead to the degeneracy which emerged in the LMOs (**Error! Reference source not found.**). Because our localized orbitals are orthonormal, small tails appear which would not be present in nonorthogonal orbitals.

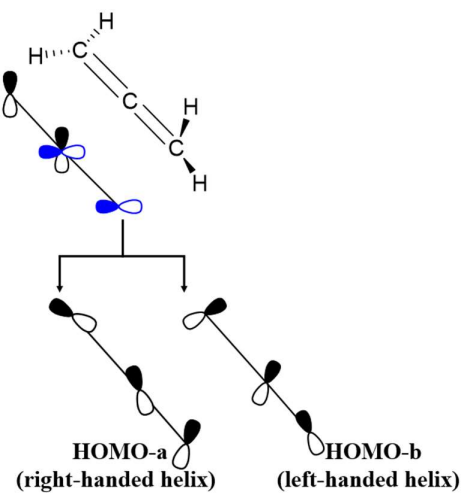

Figure 12. Standard description of $\pi$ bonding is insufficient because extended helices of specific chirality can occur.[36]

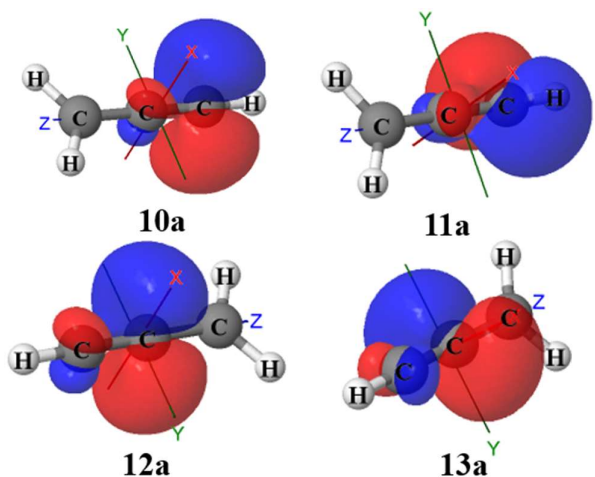

Figure 13. Helical localized molecular orbitals

## 5. Conclusions

The truncated Conjugate-Gradient Riemannian trust-region (RTR-tCG) was adapted to the localization of orthonormal orbitals. To our knowledge, this is the first use of an orbital optimization method that explicitly uses the geometry of Riemannian manifolds, through second-order expansions on charts. In particular, consideration of retractions and pullbacks allowed orthonormality to be maintained without laborious explicit orthonormalization, although occasional orthonormalizations were performed to mitigate rounding errors. The use of the truncated Conjugate-Gradient inner loop of the optimization completely avoided the need for evaluation of eigenvectors of the Hessian or solution of a set of simultaneous equations, which can be numerically problematic when the Sylvester signature has multiple large entries. In addition to localizing occupied orbitals, for which several approaches are known in the literature, the RTR-tCG algorithm was shown to be effective at obtaining orthonormal virtual orbitals, without a preliminary calculation that could introduce bias. In fact, we used the worst possible starting points in our calculations, i.e. the canonical or semi-canonical orbitals that transform as the irreducible representations of the nuclear frame, which are the results of usual quantum chemistry codes. Our algorithm is able to consider any disjoint subset of orbitals for which the underlying energy calculation is invariant, which allows us to localize a set of active orbitals in an MCSCF calculation. Although not numerically demonstrated in the current work, there is no intrinsic difficulty in MCSCF spaces with multiple partitions (as, for example, used in our macroconfiguration-based Incomplete Model Spaces). The effectiveness of the RTR-tCG algorithm was illustrated by calculations on a set of $H_{10}$ models, on *c*-$C_4H_4$, and on ground and low-lying excited states of the propargyl radical. The $H_{10}$ models probed sensitivity to the presence of diffuse orbitals, an occasional challenge to other localization schemes, and to geometries in which there are multiple orbital loci at comparable internuclear distances. The emergence of helical structures in an even [n]cumulene demonstrates the ability of RTR-tCG algorithm to provide qualitative insights for molecular orbitals in active spaces which have difficult and multiconfigurational nature. Although computational efficiency was not a goal of this pilot study, we note that each reported calculation could be performed on a single core processor in well under a minute.

In summary, it is hoped that this novel localization method will facilitate the development and use of electronic structure methods that preferentially use orthonormal orbitals in all invariant spaces. Our development considered the orthonormal group of matrices, which of course is the most straightforward of matrix groups, but this work can be expected to lead to comparable algorithms for Stiefel and Grassmann manifolds.

## Acknowledgement

The authors gratefully acknowledge the advanced cyberinfrastructure resources provided by the University of North Dakota Computational Research Center.

## Appendix. Determination of the gradient and the Hessian of $\tilde{\xi}_{FM}^m$

While calculation of the gradient and especially the Hessian on a manifold are not well-defined, their calculation on a local second-order chart can be effectively determined (cf. Eq. [6]). The (local) gradient of the fourth moment function is as,

$$\frac{\partial \tilde{\xi}_{FM}^m}{\partial \kappa_{ij}} = m \sum_p (\tilde{\mu}_4^p)^{m-1} \frac{\partial \tilde{\mu}_4^p}{\partial \kappa_{ij}} \qquad [23]$$

Note that the tilde mark (~) indicates the parametrized function. And a general expression for the (local) Hessian can be found by differentiating Eq. [23],

$$\frac{\partial^2 \tilde{\xi}_{FM}^m}{\partial \kappa_{ij} \partial \kappa_{kl}} = m \sum_p \left\{ (m-1)(\tilde{\mu}_4^p)^{m-2} \frac{\partial \tilde{\mu}_4^p}{\partial \kappa_{ij}} \frac{\partial \tilde{\mu}_4^p}{\partial \kappa_{kl}} + (\tilde{\mu}_4^p)^{m-1} \left[ \frac{\partial}{\partial \kappa_{ij}} \left( \frac{\partial \tilde{\mu}_4^p}{\partial \kappa_{kl}} \right) \right] \right\}$$

$$= m \sum_p (\tilde{\mu}_4^p)^{m-2} \left\{ (m-1) \frac{\partial \tilde{\mu}_4^p}{\partial \kappa_{ij}} \frac{\partial \tilde{\mu}_4^p}{\partial \kappa_{kl}} + \tilde{\mu}_4^p \left[ \frac{\partial}{\partial \kappa_{ij}} \left( \frac{\partial \tilde{\mu}_4^p}{\partial \kappa_{kl}} \right) \right] \right\} \qquad [24]$$

There are nine terms in Eq. [10]; their gradients and second derivatives are presented here.

**(1/9)**

$$+\langle \tilde{p} | x_t^4 | \tilde{p} \rangle \qquad [25]$$

$$\langle \tilde{p} | x_t^4 | \tilde{p} \rangle = \langle p \left| (e^{-\hat{\kappa}})^\dagger x_t^4 (e^{-\hat{\kappa}}) \right| p \rangle = \langle p | (e^{+\hat{\kappa}}) x_t^4 (e^{-\hat{\kappa}}) | p \rangle \qquad [27a]$$

$$= \langle p \left| x_t^4 + [\hat{\kappa}, x_t^4] + \frac{1}{2}[\hat{\kappa},[\hat{\kappa},x_t^4]] + \cdots \right| p \rangle$$

$$\frac{\partial}{\partial \kappa_{ij}} \langle \tilde{p} | x_t^4 | \tilde{p} \rangle = \langle p \left| \left[\frac{\partial \hat{\kappa}}{\partial \kappa_{ij}}, x_t^4\right] + \frac{1}{2}\left[\frac{\partial \hat{\kappa}}{\partial \kappa_{ij}},[\hat{\kappa},x_t^4]\right] + \frac{1}{2}\left[\hat{\kappa},\left[\frac{\partial \hat{\kappa}}{\partial \kappa_{ij}}, x_t^4\right]\right] \right. \qquad [27b]$$

$$\left. + \cdots \right| p \rangle$$

$$\left[\frac{\partial}{\partial \kappa_{ij}} \langle \tilde{p} | x_t^4 | \tilde{p} \rangle\right]_{\kappa=0} = \langle p | [(a_i^\dagger a_j - a_j^\dagger a_i), x_t^4] | p \rangle \qquad [27c]$$

$$= \langle p | (a_i^\dagger a_j - a_j^\dagger a_i) x_t^4 - x_t^4 (a_i^\dagger a_j - a_j^\dagger a_i) | p \rangle$$

$$= \delta_{ip} \langle j | x_t^4 | p \rangle - \delta_{jp} \langle i | x_t^4 | p \rangle - \langle p | x_t^4 | i \rangle \delta_{jp} + \langle p | x_t^4 | j \rangle \delta_{ip}$$

And with real orbitals,

$$\left[\frac{\partial}{\partial \kappa_{ij}}\langle \tilde{p}|x_t^4|\tilde{p}\rangle\right]_{\kappa=0} = 2\delta_{ip}\langle j|x_t^4|p\rangle - 2\delta_{jp}\langle i|x_t^4|p\rangle \qquad [27\text{d}]$$

$$\frac{\partial}{\partial \kappa_{ij}}\frac{\partial}{\partial \kappa_{kl}}\langle \tilde{p}|x_t^4|\tilde{p}\rangle \qquad [27\text{e}]$$

$$= \langle p\left|\frac{1}{2}\left[\frac{\partial \hat{\kappa}}{\partial \kappa_{ij}},\left[\frac{\partial \hat{\kappa}}{\partial \kappa_{kl}},x_t^4\right]\right] + \frac{1}{2}\left[\frac{\partial \hat{\kappa}}{\partial \kappa_{kl}},\left[\frac{\partial \hat{\kappa}}{\partial \kappa_{ij}},x_t^4\right]\right] + \cdots\right|p\rangle$$

$$\left[\frac{\partial}{\partial \kappa_{ij}}\frac{\partial}{\partial \kappa_{kl}}\langle \tilde{p}|x_t^4|\tilde{p}\rangle\right]_{\kappa=0} \qquad [27\text{f}]$$

$$= \frac{1}{2}\langle p\left|\left[(a_i^\dagger a_j - a_j^\dagger a_i),\left[(a_k^\dagger a_l - a_l^\dagger a_k),x_t^4\right]\right]\right|p\rangle$$

$$+ \frac{1}{2}\langle p\left|\left[(a_k^\dagger a_l - a_l^\dagger a_k),\left[(a_i^\dagger a_j - a_j^\dagger a_i),x_t^4\right]\right]\right|p\rangle$$

$$\left[\frac{\partial}{\partial \kappa_{ij}}\frac{\partial}{\partial \kappa_{kl}}\langle \tilde{p}|x_t^4|\tilde{p}\rangle\right]_{\kappa=0} \qquad [27\text{g}]$$

$$= \frac{1}{2}\langle p|(a_i^\dagger a_j - a_j^\dagger a_i)[(a_k^\dagger a_l - a_l^\dagger a_k),x_t^4]$$

$$- [(a_k^\dagger a_l - a_l^\dagger a_k),x_t^4](a_i^\dagger a_j - a_j^\dagger a_i)|p\rangle$$

$$+ \frac{1}{2}\langle p|(a_k^\dagger a_l - a_l^\dagger a_k)[(a_i^\dagger a_j - a_j^\dagger a_i),x_t^4]$$

$$- [(a_i^\dagger a_j - a_j^\dagger a_i),x_t^4](a_k^\dagger a_l - a_l^\dagger a_k)|p\rangle$$

$$\left[\frac{\partial}{\partial \kappa_{ij}}\frac{\partial}{\partial \kappa_{kl}}\langle \tilde{p}|x_t^4|\tilde{p}\rangle\right]_{\kappa=0} \qquad [27\text{h}]$$

$$= \frac{1}{2}\delta_{ip}\langle j|[(a_k^\dagger a_l - a_l^\dagger a_k),x_t^4]|p\rangle$$

$$- \frac{1}{2}\delta_{jp}\langle i|[(a_k^\dagger a_l - a_l^\dagger a_k),x_t^4]|p\rangle$$

$$- \frac{1}{2}\langle p|[(a_k^\dagger a_l - a_l^\dagger a_k),x_t^4]|i\rangle\delta_{jp}$$

$$+ \frac{1}{2}\langle p|[(a_k^\dagger a_l - a_l^\dagger a_k),x_t^4]|j\rangle\delta_{ip}$$

$$+ \frac{1}{2}\delta_{kp}\langle l|[(a_i^\dagger a_j - a_j^\dagger a_i),x_t^4]|p\rangle$$

$$- \frac{1}{2}\delta_{lp}\langle k|[(a_i^\dagger a_j - a_j^\dagger a_i),x_t^4]|p\rangle$$

$$- \frac{1}{2}\langle p|[(a_i^\dagger a_j - a_j^\dagger a_i),x_t^4]|k\rangle\delta_{lp}$$

$$+ \frac{1}{2}\langle p|[(a_i^\dagger a_j - a_j^\dagger a_i),x_t^4]|l\rangle\delta_{kp}$$

$$\left[\frac{\partial}{\partial \kappa_{ij}}\frac{\partial}{\partial \kappa_{kl}}\langle \tilde{p}|x_t^4|\tilde{p}\rangle\right]_{\kappa=0} \quad [27\text{i}]$$

$$= \frac{1}{2}\delta_{ip}\langle j|(a_k^\dagger a_l - a_l^\dagger a_k)x_t^4 - x_t^4(a_k^\dagger a_l - a_l^\dagger a_k)|p\rangle$$
$$- \frac{1}{2}\delta_{jp}\langle i|(a_k^\dagger a_l - a_l^\dagger a_k)x_t^4 - x_t^4(a_k^\dagger a_l - a_l^\dagger a_k)|p\rangle$$
$$- \frac{1}{2}\langle p|(a_k^\dagger a_l - a_l^\dagger a_k)x_t^4 - x_t^4(a_k^\dagger a_l - a_l^\dagger a_k)|i\rangle\delta_{jp}$$
$$+ \frac{1}{2}\langle p|(a_k^\dagger a_l - a_l^\dagger a_k)x_t^4 - x_t^4(a_k^\dagger a_l - a_l^\dagger a_k)|j\rangle\delta_{ip}$$
$$+ \frac{1}{2}\delta_{kp}\langle l|(a_i^\dagger a_j - a_j^\dagger a_i)x_t^4 - x_t^4(a_i^\dagger a_j - a_j^\dagger a_i)|p\rangle$$
$$- \frac{1}{2}\delta_{lp}\langle k|(a_i^\dagger a_j - a_j^\dagger a_i)x_t^4 - x_t^4(a_i^\dagger a_j - a_j^\dagger a_i)|p\rangle$$
$$- \frac{1}{2}\langle p|(a_i^\dagger a_j - a_j^\dagger a_i)x_t^4 - x_t^4(a_i^\dagger a_j - a_j^\dagger a_i)|k\rangle\delta_{lp}$$
$$+ \frac{1}{2}\langle p|(a_i^\dagger a_j - a_j^\dagger a_i)x_t^4 - x_t^4(a_i^\dagger a_j - a_j^\dagger a_i)|l\rangle\delta_{kp}$$

And with real orbitals,

$$\left[\frac{\partial}{\partial \kappa_{ij}}\frac{\partial}{\partial \kappa_{kl}}\langle \tilde{p}|x_t^4|\tilde{p}\rangle\right]_{\kappa=0} \quad [27\text{j}]$$
$$= +\delta_{ip}\langle j|(a_k^\dagger a_l - a_l^\dagger a_k)x_t^4 - x_t^4(a_k^\dagger a_l - a_l^\dagger a_k)|p\rangle$$
$$- \delta_{jp}\langle i|(a_k^\dagger a_l - a_l^\dagger a_k)x_t^4 - x_t^4(a_k^\dagger a_l - a_l^\dagger a_k)|p\rangle$$
$$+ \delta_{kp}\langle l|(a_i^\dagger a_j - a_j^\dagger a_i)x_t^4 - x_t^4(a_i^\dagger a_j - a_j^\dagger a_i)|p\rangle$$
$$- \delta_{lp}\langle k|(a_i^\dagger a_j - a_j^\dagger a_i)x_t^4 - x_t^4(a_i^\dagger a_j - a_j^\dagger a_i)|p\rangle$$

$$\left[\frac{\partial}{\partial \kappa_{ij}}\frac{\partial}{\partial \kappa_{kl}}\langle \tilde{p}|x_t^4|\tilde{p}\rangle\right]_{\kappa=0} \quad [27\text{k}]$$
$$= +\delta_{ip}\delta_{jk}\langle l|x_t^4|p\rangle - \delta_{ip}\delta_{jl}\langle k|x_t^4|p\rangle - \delta_{ip}\langle j|x_t^4|k\rangle\delta_{lp}$$
$$+ \delta_{ip}\langle j|x_t^4|l\rangle\delta_{kp} - \delta_{jp}\delta_{ik}\langle l|x_t^4|p\rangle + \delta_{jp}\delta_{il}\langle k|x_t^4|p\rangle$$
$$+ \delta_{jp}\langle i|x_t^4|k\rangle\delta_{lp} - \delta_{jp}\langle i|x_t^4|l\rangle\delta_{kp} + \delta_{kp}\delta_{il}\langle j|x_t^4|p\rangle$$
$$- \delta_{kp}\delta_{lj}\langle i|x_t^4|p\rangle - \delta_{kp}\langle l|x_t^4|i\rangle\delta_{jp} + \delta_{kp}\langle l|x_t^4|j\rangle\delta_{ip}$$
$$- \delta_{lp}\delta_{ik}\langle j|x_t^4|p\rangle + \delta_{lp}\delta_{jk}\langle i|x_t^4|p\rangle + \delta_{lp}\langle k|x_t^4|i\rangle\delta_{jp}$$
$$- \delta_{lp}\langle k|x_t^4|j\rangle\delta_{ip}$$

$$\left[\frac{\partial}{\partial \kappa_{ij}}\frac{\partial}{\partial \kappa_{kl}}\langle \tilde{p}|x_t^4|\tilde{p}\rangle\right]_{\kappa=0} \quad [27\text{l}]$$
$$= +\delta_{ip}\{\delta_{jk}\langle l|x_t^4|p\rangle - \delta_{jl}\langle k|x_t^4|p\rangle - 2\langle j|x_t^4|k\rangle\delta_{lp}$$
$$+ 2\langle j|x_t^4|l\rangle\delta_{kp}\}$$
$$- \delta_{jp}\{\delta_{ik}\langle l|x_t^4|p\rangle - \delta_{il}\langle k|x_t^4|p\rangle - 2\langle i|x_t^4|k\rangle\delta_{lp}$$
$$+ 2\langle i|x_t^4|l\rangle\delta_{kp}\} + \delta_{kp}\{\delta_{il}\langle j|x_t^4|p\rangle - \delta_{lj}\langle i|x_t^4|p\rangle\}$$
$$- \delta_{lp}\{\delta_{ik}\langle j|x_t^4|p\rangle - \delta_{jk}\langle i|x_t^4|p\rangle\}$$

This simplifies essentially for the diagonal ($i = k, j = l, \langle i|x_t^4|j\rangle = 0$),

$$\left[\frac{\partial}{\partial\kappa_{ij}}\frac{\partial}{\partial\kappa_{ij}}\langle\tilde{p}|x_t^4|\tilde{p}\rangle\right]_{\kappa=0} \qquad [27\text{m}]$$
$$= +\delta_{ip}\{-2\langle i|x_t^4|p\rangle + 2\langle j|x_t^4|j\rangle\}$$
$$- \delta_{jp}\{+2\langle j|x_t^4|p\rangle - 2\langle i|x_t^4|i\rangle\}$$

This diagonal second derivative can be written more compactly as,

$$\left[\frac{\partial}{\partial\kappa_{ij}}\frac{\partial}{\partial\kappa_{ij}}\langle\tilde{p}|x_t^4|\tilde{p}\rangle\right]_{\kappa=0} \qquad [27\text{n}]$$
$$= +\delta_{ip}\{+2\langle j|x_t^4|j\rangle - 2\langle p|x_t^4|p\rangle\}$$
$$+ \delta_{jp}\{+2\langle i|x_t^4|i\rangle - 2\langle p|x_t^4|p\rangle\}$$

**(2/9)**

$$-4\langle\tilde{p}|x_t^3|\tilde{p}\rangle\langle\tilde{p}|x_t|\tilde{p}\rangle \qquad [26]$$

$$\frac{\partial}{\partial\kappa_{ij}}\langle\tilde{p}|x_t^3|\tilde{p}\rangle\langle\tilde{p}|x_t|\tilde{p}\rangle \qquad [28\text{a}]$$
$$= \left\{\frac{\partial}{\partial\kappa_{ij}}\langle\tilde{p}|x_t^3|\tilde{p}\rangle\right\}\langle\tilde{p}|x_t|\tilde{p}\rangle + \langle\tilde{p}|x_t^3|\tilde{p}\rangle\left\{\frac{\partial}{\partial\kappa_{ij}}\langle\tilde{p}|x_t|\tilde{p}\rangle\right\}$$

$$\left[\frac{\partial}{\partial\kappa_{ij}}\langle\tilde{p}|x_t^3|\tilde{p}\rangle\langle\tilde{p}|x_t|\tilde{p}\rangle\right]_{\kappa=0} \qquad [28\text{b}]$$
$$= \{2\delta_{ip}\langle j|x_t^3|p\rangle - 2\delta_{jp}\langle i|x_t^3|p\rangle\}\langle p|x_t|p\rangle$$
$$+ \langle p|x_t^3|p\rangle\{2\delta_{ip}\langle j|x_t|p\rangle - 2\delta_{jp}\langle i|x_t|p\rangle\}$$

$$\frac{\partial}{\partial\kappa_{ij}}\frac{\partial}{\partial\kappa_{kl}}\langle\tilde{p}|x_t^3|\tilde{p}\rangle\langle\tilde{p}|x_t|\tilde{p}\rangle \qquad [28\text{c}]$$
$$= \left\{\frac{\partial}{\partial\kappa_{ij}}\frac{\partial}{\partial\kappa_{kl}}\langle\tilde{p}|x_t^3|\tilde{p}\rangle\right\}\langle\tilde{p}|x_t|\tilde{p}\rangle$$
$$+ \left\{\frac{\partial}{\partial\kappa_{ij}}\langle\tilde{p}|x_t^3|\tilde{p}\rangle\right\}\left\{\frac{\partial}{\partial\kappa_{kl}}\langle\tilde{p}|x_t|\tilde{p}\rangle\right\}$$
$$+ \left\{\frac{\partial}{\partial\kappa_{kl}}\langle\tilde{p}|x_t^3|\tilde{p}\rangle\right\}\left\{\frac{\partial}{\partial\kappa_{ij}}\langle\tilde{p}|x_t|\tilde{p}\rangle\right\}$$
$$+ \langle\tilde{p}|x_t^3|\tilde{p}\rangle\left\{\frac{\partial}{\partial\kappa_{ij}}\frac{\partial}{\partial\kappa_{kl}}\langle\tilde{p}|x_t|\tilde{p}\rangle\right\}$$

$$\left[\frac{\partial}{\partial\kappa_{ij}}\frac{\partial}{\partial\kappa_{kl}}\langle\tilde{p}|x_t^3|\tilde{p}\rangle\langle\tilde{p}|x_t|\tilde{p}\rangle\right]_{\kappa=0} \qquad [28\text{d}]$$
$$= \begin{cases} +\delta_{ip}\{\delta_{jk}\langle l|x_t^3|p\rangle - \delta_{jl}\langle k|x_t^3|p\rangle - 2\langle j|x_t^3|k\rangle\delta_{lp} + 2\langle j|x_t^3|l\rangle\delta_{kp}\} \\ -\delta_{jp}\{\delta_{ik}\langle l|x_t^3|p\rangle - \delta_{il}\langle k|x_t^3|p\rangle - 2\langle i|x_t^3|k\rangle\delta_{lp} + 2\langle i|x_t^3|l\rangle\delta_{kp}\} \\ +\delta_{kp}\{\delta_{il}\langle j|x_t^3|p\rangle - \delta_{lj}\langle i|x_t^3|p\rangle\} - \delta_{lp}\{\delta_{ik}\langle j|x_t^3|p\rangle - \delta_{jk}\langle i|x_t^3|p\rangle\} \end{cases} \times$$
$$\langle p|x_t|p\rangle + \{2\delta_{ip}\langle j|x_t^3|p\rangle - 2\delta_{jp}\langle i|x_t^3|p\rangle\}\{2\delta_{kp}\langle l|x_t|p\rangle - 2\delta_{lp}\langle k|x_t|p\rangle\}$$
$$+ \{2\delta_{kp}\langle l|x_t^3|p\rangle - 2\delta_{lp}\langle k|x_t^3|p\rangle\}\{2\delta_{ip}\langle j|x_t|p\rangle$$
$$- 2\delta_{jp}\langle i|x_t|p\rangle\} + \langle p|x_t^3|p\rangle \times$$

$$\left\{\begin{array}{l}+\delta_{ip}\{\delta_{jk}\langle l|x_t|p\rangle-\delta_{jl}\langle k|x_t|p\rangle-2\langle j|x_t|k\rangle\delta_{lp}+2\langle j|x_t|l\rangle\delta_{kp}\}\\-\delta_{jp}\{\delta_{ik}\langle l|x_t|p\rangle-\delta_{il}\langle k|x_t|p\rangle-2\langle i|x_t|k\rangle\delta_{lp}+2\langle i|x_t|l\rangle\delta_{kp}\}\\+\delta_{kp}\{\delta_{il}\langle j|x_t|p\rangle-\delta_{lj}\langle i|x_t|p\rangle\}-\delta_{lp}\{\delta_{ik}\langle j|x_t|p\rangle-\delta_{jk}\langle i|x_t|p\rangle\}\end{array}\right\}$$

**(3/9)**

$$+6\langle\tilde{p}|x_t^2|\tilde{p}\rangle\langle\tilde{p}|x_t|\tilde{p}\rangle^2 \qquad [27]$$

$$\frac{\partial}{\partial\kappa_{ij}}\langle\tilde{p}|x_t^2|\tilde{p}\rangle\langle\tilde{p}|x_t|\tilde{p}\rangle^2 \qquad [29a]$$

$$=\left\{\frac{\partial}{\partial\kappa_{ij}}\langle\tilde{p}|x_t^2|\tilde{p}\rangle\right\}\langle\tilde{p}|x_t|\tilde{p}\rangle^2$$

$$+\langle\tilde{p}|x_t^2|\tilde{p}\rangle\left\{2\langle\tilde{p}|x_t|\tilde{p}\rangle\left\{\frac{\partial}{\partial\kappa_{ij}}\langle\tilde{p}|x_t|\tilde{p}\rangle\right\}\right\}$$

$$\left[\frac{\partial}{\partial\kappa_{ij}}\langle\tilde{p}|x_t^2|\tilde{p}\rangle\langle\tilde{p}|x_t|\tilde{p}\rangle^2\right]_{\kappa=0} \qquad [29b]$$

$$=\{2\delta_{ip}\langle j|x_t^2|p\rangle-2\delta_{jp}\langle i|x_t^2|p\rangle\}\langle p|x_t|p\rangle^2$$

$$+\langle p|x_t^2|p\rangle\left\{2\langle p|x_t|p\rangle\{2\delta_{ip}\langle j|x_t|p\rangle-2\delta_{jp}\langle i|x_t|p\rangle\}\right\}$$

$$\frac{\partial}{\partial\kappa_{ij}}\frac{\partial}{\partial\kappa_{kl}}\langle\tilde{p}|x_t^2|\tilde{p}\rangle\langle\tilde{p}|x_t|\tilde{p}\rangle^2 \qquad [29c]$$

$$=\left\{\frac{\partial}{\partial\kappa_{ij}}\frac{\partial}{\partial\kappa_{kl}}\langle\tilde{p}|x_t^2|\tilde{p}\rangle\right\}\langle\tilde{p}|x_t|\tilde{p}\rangle^2$$

$$+\left\{\frac{\partial}{\partial\kappa_{ij}}\langle\tilde{p}|x_t^2|\tilde{p}\rangle\right\}\left\{2\langle\tilde{p}|x_t|\tilde{p}\rangle\left\{\frac{\partial}{\partial\kappa_{kl}}\langle\tilde{p}|x_t|\tilde{p}\rangle\right\}\right\}$$

$$+\left\{\frac{\partial}{\partial\kappa_{kl}}\langle\tilde{p}|x_t^2|\tilde{p}\rangle\right\}\left\{2\langle\tilde{p}|x_t|\tilde{p}\rangle\left\{\frac{\partial}{\partial\kappa_{ij}}\langle\tilde{p}|x_t|\tilde{p}\rangle\right\}\right\}$$

$$+\langle\tilde{p}|x_t^2|\tilde{p}\rangle\left\{2\frac{\partial}{\partial\kappa_{kl}}\langle\tilde{p}|x_t|\tilde{p}\rangle\right\}\left\{\frac{\partial}{\partial\kappa_{ij}}\langle\tilde{p}|x_t|\tilde{p}\rangle\right\}$$

$$+\langle\tilde{p}|x_t^2|\tilde{p}\rangle\left\{2\langle\tilde{p}|x_t|\tilde{p}\rangle\left\{\frac{\partial}{\partial\kappa_{ij}}\frac{\partial}{\partial\kappa_{kl}}\langle\tilde{p}|x_t|\tilde{p}\rangle\right\}\right\}$$

$$\left[\frac{\partial}{\partial\kappa_{ij}}\frac{\partial}{\partial\kappa_{kl}}\langle\tilde{p}|x_t^2|\tilde{p}\rangle\langle\tilde{p}|x_t|\tilde{p}\rangle^2\right]_{\kappa=0} \qquad [29d]$$

$$=\left\{\begin{array}{l}+\delta_{ip}\{\delta_{jk}\langle l|x_t^2|p\rangle-\delta_{jl}\langle k|x_t^2|p\rangle-2\langle j|x_t^2|k\rangle\delta_{lp}+2\langle j|x_t^2|l\rangle\delta_{kp}\}\\-\delta_{jp}\{\delta_{ik}\langle l|x_t^2|p\rangle-\delta_{il}\langle k|x_t^2|p\rangle-2\langle i|x_t^2|k\rangle\delta_{lp}+2\langle i|x_t^2|l\rangle\delta_{kp}\}\\+\delta_{kp}\{\delta_{il}\langle j|x_t^2|p\rangle-\delta_{lj}\langle i|x_t^2|p\rangle\}-\delta_{lp}\{\delta_{ik}\langle j|x_t^2|p\rangle-\delta_{jk}\langle i|x_t^2|p\rangle\}\end{array}\right\}\times$$

$$\langle p|x_t|p\rangle^2 + \{2\delta_{ip}\langle j|x_t^2|p\rangle$$
$$- 2\delta_{jp}\langle i|x_t^2|p\rangle\}\{2\langle p|x_t|p\rangle\{2\delta_{kp}\langle l|x_t|p\rangle - 2\delta_{lp}\langle k|x_t|p\rangle\}\}$$
$$+ \{2\delta_{kp}\langle l|x_t^2|p\rangle$$
$$- 2\delta_{lp}\langle k|x_t^2|p\rangle\}\{2\langle p|x_t|p\rangle\{2\delta_{ip}\langle j|x_t|p\rangle - 2\delta_{jp}\langle i|x_t|p\rangle\}\}$$
$$+ \langle p|x_t^2|p\rangle\{2\{2\delta_{kp}\langle l|x_t|p\rangle - 2\delta_{lp}\langle k|x_t|p\rangle\}\}\{2\delta_{ip}\langle j|x_t|p\rangle$$
$$- 2\delta_{jp}\langle i|x_t|p\rangle\} + \langle p|x_t^2|p\rangle \times$$
$$\left\{ \begin{pmatrix} 2\langle p|x_t|p\rangle \\ +\delta_{ip}\{\delta_{jk}\langle l|x_t|p\rangle - \delta_{jl}\langle k|x_t|p\rangle - 2\langle j|x_t|k\rangle\delta_{lp} + 2\langle j|x_t|l\rangle\delta_{kp}\} \\ -\delta_{jp}\{\delta_{ik}\langle l|x_t|p\rangle - \delta_{il}\langle k|x_t|p\rangle - 2\langle i|x_t|k\rangle\delta_{lp} + 2\langle i|x_t|l\rangle\delta_{kp}\} \\ +\delta_{kp}\{\delta_{il}\langle j|x_t|p\rangle - \delta_{lj}\langle i|x_t|p\rangle\} - \delta_{lp}\{\delta_{ik}\langle j|x_t|p\rangle - \delta_{jk}\langle i|x_t|p\rangle\} \end{pmatrix} \right\}$$

**(4/9)**

$$-3\langle\tilde{p}|x_t|\tilde{p}\rangle^4 \tag{28}$$

$$\frac{\partial}{\partial\kappa_{ij}}\langle\tilde{p}|x_t|\tilde{p}\rangle^4 = 4\langle\tilde{p}|x_t|\tilde{p}\rangle^3\left\{\frac{\partial}{\partial\kappa_{ij}}\langle\tilde{p}|x_t|\tilde{p}\rangle\right\} \tag{30a}$$

$$\left[\frac{\partial}{\partial\kappa_{ij}}\langle\tilde{p}|x_t|\tilde{p}\rangle^4\right]_{\kappa=0} = 4\langle p|x_t|p\rangle^3\{2\delta_{ip}\langle j|x_t|p\rangle - 2\delta_{jp}\langle i|x_t|p\rangle\} \tag{30b}$$

$$\frac{\partial}{\partial\kappa_{ij}}\frac{\partial}{\partial\kappa_{kl}}\langle\tilde{p}|x_t|\tilde{p}\rangle^4 \tag{30c}$$

$$= 4\left\{3\langle\tilde{p}|x_t|\tilde{p}\rangle^2\left\{\frac{\partial}{\partial\kappa_{kl}}\langle\tilde{p}|x_t|\tilde{p}\rangle\right\}\right\}\left\{\frac{\partial}{\partial\kappa_{ij}}\langle\tilde{p}|x_t|\tilde{p}\rangle\right\}$$
$$+ 4\langle\tilde{p}|x_t|\tilde{p}\rangle^3\left\{\frac{\partial}{\partial\kappa_{ij}}\frac{\partial}{\partial\kappa_{kl}}\langle\tilde{p}|x_t|\tilde{p}\rangle\right\}$$

$$\left[\frac{\partial}{\partial\kappa_{ij}}\frac{\partial}{\partial\kappa_{kl}}\langle\tilde{p}|x_t|\tilde{p}\rangle^4\right]_{\kappa=0} \tag{30d}$$
$$= 12\left\{\langle p|x_t|p\rangle^2\{2\delta_{kp}\langle l|x_t|p\rangle - 2\delta_{lp}\langle k|x_t|p\rangle\}\right\}\{2\delta_{ip}\langle j|x_t|p\rangle$$
$$- 2\delta_{jp}\langle i|x_t|p\rangle\} + 4\langle p|x_t|p\rangle^3 \times$$
$$\begin{pmatrix} +\delta_{ip}\{\delta_{jk}\langle l|x_t|p\rangle - \delta_{jl}\langle k|x_t|p\rangle - 2\langle j|x_t|k\rangle\delta_{lp} + 2\langle j|x_t|l\rangle\delta_{kp}\} \\ -\delta_{jp}\{\delta_{ik}\langle l|x_t|p\rangle - \delta_{il}\langle k|x_t|p\rangle - 2\langle i|x_t|k\rangle\delta_{lp} + 2\langle i|x_t|l\rangle\delta_{kp}\} \\ +\delta_{kp}\{\delta_{il}\langle j|x_t|p\rangle - \delta_{lj}\langle i|x_t|p\rangle\} - \delta_{lp}\{\delta_{ik}\langle j|x_t|p\rangle - \delta_{jk}\langle i|x_t|p\rangle\} \end{pmatrix}$$

**(5/9)**

$$+2\langle\tilde{p}|x_t^2 x_u^2|\tilde{p}\rangle \tag{29}$$

$$\langle\tilde{p}|x_t^2 x_u^2|\tilde{p}\rangle = \langle p\left|x_t^2 x_u^2 + [\hat{\kappa}, x_t^2 x_u^2] + \frac{1}{2}\left[\hat{\kappa},[\hat{\kappa}, x_t^2 x_u^2]\right] + \cdots\right|p\rangle \tag{31a}$$

$$\frac{\partial}{\partial \kappa_{ij}} \langle \tilde{p}|x_t^2 x_u^2|\tilde{p}\rangle \qquad [31b]$$

$$= \langle p \left| \left[ \frac{\partial \hat{\kappa}}{\partial \kappa_{ij}}, x_t^2 x_u^2 \right] + \frac{1}{2} \left[ \frac{\partial \hat{\kappa}}{\partial \kappa_{ij}}, [\hat{\kappa}, x_t^2 x_u^2] \right] \right.$$

$$\left. + \frac{1}{2} \left[ \hat{\kappa}, \left[ \frac{\partial \hat{\kappa}}{\partial \kappa_{ij}}, x_t^2 x_u^2 \right] \right] + \cdots \right| p \rangle$$

$$\left[ \frac{\partial}{\partial \kappa_{ij}} \langle \tilde{p}|x_t^2 x_u^2|\tilde{p}\rangle \right]_{\kappa=0} = 2\delta_{ip}\langle j|x_t^2 x_u^2|p\rangle - 2\delta_{jp}\langle i|x_t^2 x_u^2|p\rangle \qquad [31c]$$

$$\frac{\partial}{\partial \kappa_{ij}} \frac{\partial}{\partial \kappa_{kl}} \langle \tilde{p}|x_t^2 x_u^2|\tilde{p}\rangle \qquad [31d]$$

$$= \langle p \left| \frac{1}{2} \left[ \frac{\partial \hat{\kappa}}{\partial \kappa_{ij}}, \left[ \frac{\partial \hat{\kappa}}{\partial \kappa_{kl}}, x_t^2 x_u^2 \right] \right] + \frac{1}{2} \left[ \frac{\partial \hat{\kappa}}{\partial \kappa_{kl}}, \left[ \frac{\partial \hat{\kappa}}{\partial \kappa_{ij}}, x_t^2 x_u^2 \right] \right] \right.$$

$$\left. + \cdots \right| p \rangle$$

$$\left[ \frac{\partial}{\partial \kappa_{ij}} \frac{\partial}{\partial \kappa_{kl}} \langle \tilde{p}|x_t^2 x_u^2|\tilde{p}\rangle \right]_{\kappa=0} \qquad [31e]$$

$$= +\delta_{ip}\{\delta_{jk}\langle l|x_t^2 x_u^2|p\rangle - \delta_{jl}\langle k|x_t^2 x_u^2|p\rangle - 2\langle j|x_t^2 x_u^2|k\rangle\delta_{lp}$$
$$+ 2\langle j|x_t^2 x_u^2|l\rangle\delta_{kp}\}$$
$$- \delta_{jp}\{\delta_{ik}\langle l|x_t^2 x_u^2|p\rangle - \delta_{il}\langle k|x_t^2 x_u^2|p\rangle - 2\langle i|x_t^2 x_u^2|k\rangle\delta_{lp}$$
$$+ 2\langle i|x_t^2 x_u^2|l\rangle\delta_{kp}\} + \delta_{kp}\{\delta_{il}\langle j|x_t^2 x_u^2|p\rangle - \delta_{lj}\langle i|x_t^2 x_u^2|p\rangle\}$$
$$- \delta_{lp}\{\delta_{ik}\langle j|x_t^2 x_u^2|p\rangle - \delta_{jk}\langle i|x_t^2 x_u^2|p\rangle\}$$

**(6/9)**

$$-4(1 + \hat{P}_{tu})\langle \tilde{p}|x_t^2 x_u|\tilde{p}\rangle\langle \tilde{p}|x_u|\tilde{p}\rangle \qquad [30]$$

$$\frac{\partial}{\partial \kappa_{ij}} \langle \tilde{p}|x_t^2 x_u|\tilde{p}\rangle\langle \tilde{p}|x_u|\tilde{p}\rangle \qquad [32a]$$

$$= \left\{ \frac{\partial}{\partial \kappa_{ij}} \langle \tilde{p}|x_t^2 x_u|\tilde{p}\rangle \right\} \langle \tilde{p}|x_u|\tilde{p}\rangle + \langle \tilde{p}|x_t^2 x_u|\tilde{p}\rangle \left\{ \frac{\partial}{\partial \kappa_{ij}} \langle \tilde{p}|x_u|\tilde{p}\rangle \right\}$$

$$\left[ \frac{\partial}{\partial \kappa_{ij}} \langle \tilde{p}|x_t^2 x_u|\tilde{p}\rangle\langle \tilde{p}|x_u|\tilde{p}\rangle \right]_{\kappa=0} \qquad [32b]$$

$$= \{2\delta_{ip}\langle j|x_t^2 x_u|p\rangle - 2\delta_{jp}\langle i|x_t^2 x_u|p\rangle\}\langle p|x_u|p\rangle$$
$$+ \langle p|x_t^2 x_u|p\rangle\{2\delta_{ip}\langle j|x_u|p\rangle - 2\delta_{jp}\langle i|x_u|p\rangle\}$$

$$\frac{\partial}{\partial \kappa_{ij}} \frac{\partial}{\partial \kappa_{kl}} \langle \tilde{p}|x_t^2 x_u|\tilde{p}\rangle\langle \tilde{p}|x_u|\tilde{p}\rangle \qquad [32c]$$

$$= \left\{\frac{\partial}{\partial \kappa_{ij}} \frac{\partial}{\partial \kappa_{kl}} \langle \tilde{p}|x_t^2 x_u|\tilde{p}\rangle\right\} \langle \tilde{p}|x_u|\tilde{p}\rangle$$

$$+ \left\{\frac{\partial}{\partial \kappa_{ij}} \langle \tilde{p}|x_t^2 x_u|\tilde{p}\rangle\right\}\left\{\frac{\partial}{\partial \kappa_{kl}} \langle \tilde{p}|x_u|\tilde{p}\rangle\right\}$$

$$+ \left\{\frac{\partial}{\partial \kappa_{kl}} \langle \tilde{p}|x_t^2 x_u|\tilde{p}\rangle\right\}\left\{\frac{\partial}{\partial \kappa_{ij}} \langle \tilde{p}|x_u|\tilde{p}\rangle\right\}$$

$$+ \langle \tilde{p}|x_t^2 x_u|\tilde{p}\rangle \left\{\frac{\partial}{\partial \kappa_{ij}} \frac{\partial}{\partial \kappa_{kl}} \langle \tilde{p}|x_u|\tilde{p}\rangle\right\}$$

$$\left[\frac{\partial}{\partial \kappa_{ij}} \frac{\partial}{\partial \kappa_{kl}} \langle \tilde{p}|x_t^2 x_u|\tilde{p}\rangle\langle \tilde{p}|x_u|\tilde{p}\rangle\right]_{\kappa=0} \qquad [32d]$$

$$= \begin{cases} +\delta_{ip} \begin{Bmatrix} \delta_{jk}\langle l|x_t^2 x_u|p\rangle - \delta_{jl}\langle k|x_t^2 x_u|p\rangle \\ -2\langle j|x_t^2 x_u|k\rangle\delta_{lp} + 2\langle j|x_t^2 x_u|l\rangle\delta_{kp} \end{Bmatrix} \\ -\delta_{jp} \begin{Bmatrix} \delta_{ik}\langle l|x_t^2 x_u|p\rangle - \delta_{il}\langle k|x_t^2 x_u|p\rangle \\ -2\langle i|x_t^2 x_u|k\rangle\delta_{lp} + 2\langle i|x_t^2 x_u|l\rangle\delta_{kp} \end{Bmatrix} \\ +\delta_{kp}\{\delta_{il}\langle j|x_t^2 x_u|p\rangle - \delta_{lj}\langle i|x_t^2 x_u|p\rangle\} \\ -\delta_{lp}\{\delta_{ik}\langle j|x_t^2 x_u|p\rangle - \delta_{jk}\langle i|x_t^2 x_u|p\rangle\} \end{cases} \langle p|x_u|p\rangle$$

$$+ \{2\delta_{ip}\langle j|x_t^2 x_u|p\rangle - 2\delta_{jp}\langle i|x_t^2 x_u|p\rangle\}\{2\delta_{kp}\langle l|x_u|p\rangle - 2\delta_{lp}\langle k|x_u|p\rangle\}$$

$$+ \{2\delta_{kp}\langle l|x_t^2 x_u|p\rangle - 2\delta_{lp}\langle k|x_t^2 x_u|p\rangle\}\{2\delta_{ip}\langle j|x_u|p\rangle - 2\delta_{jp}\langle i|x_u|p\rangle\}$$

$$+ \langle p|x_t^2 x_u|p\rangle \begin{cases} +\delta_{ip} \begin{Bmatrix} \delta_{jk}\langle l|x_u|p\rangle - \delta_{jl}\langle k|x_u|p\rangle \\ -2\langle j|x_u|k\rangle\delta_{lp} + 2\langle j|x_u|l\rangle\delta_{kp} \end{Bmatrix} \\ -\delta_{jp} \begin{Bmatrix} \delta_{ik}\langle l|x_u|p\rangle - \delta_{il}\langle k|x_u|p\rangle \\ -2\langle i|x_u|k\rangle\delta_{lp} + 2\langle i|x_u|l\rangle\delta_{kp} \end{Bmatrix} \\ +\delta_{kp}\{\delta_{il}\langle j|x_u|p\rangle - \delta_{lj}\langle i|x_u|p\rangle\} \\ -\delta_{lp}\{\delta_{ik}\langle j|x_u|p\rangle - \delta_{jk}\langle i|x_u|p\rangle\} \end{cases}$$

**(7/9)**

$$+2(1+\hat{P}_{tu})\langle \tilde{p}|x_t^2|\tilde{p}\rangle\langle \tilde{p}|x_u|\tilde{p}\rangle^2 \qquad [31]$$

$$\frac{\partial}{\partial \kappa_{ij}} \langle \tilde{p}|x_t^2|\tilde{p}\rangle\langle \tilde{p}|x_u|\tilde{p}\rangle^2 \qquad [33a]$$

$$= \left\{\frac{\partial}{\partial \kappa_{ij}} \langle \tilde{p}|x_t^2|\tilde{p}\rangle\right\} \langle \tilde{p}|x_u|\tilde{p}\rangle^2$$

$$+ \langle \tilde{p}|x_t^2|\tilde{p}\rangle \left\{2\langle \tilde{p}|x_u|\tilde{p}\rangle \frac{\partial}{\partial \kappa_{ij}} \langle \tilde{p}|x_u|\tilde{p}\rangle\right\}$$

$$\left[\frac{\partial}{\partial \kappa_{ij}} \langle \tilde{p}|x_t^2|\tilde{p}\rangle \langle \tilde{p}|x_u|\tilde{p}\rangle^2 \right]_{\kappa=0} \tag{33b}$$
$$= \{2\delta_{ip}\langle j|x_t^2|p\rangle - 2\delta_{jp}\langle i|x_t^2|p\rangle\}\langle p|x_u|p\rangle^2$$
$$+ 2\langle p|x_t^2|p\rangle \langle p|x_u|p\rangle\{2\delta_{ip}\langle j|x_u|p\rangle - 2\delta_{jp}\langle i|x_u|p\rangle\}$$

$$\frac{\partial}{\partial \kappa_{ij}} \frac{\partial}{\partial \kappa_{kl}} \langle \tilde{p}|x_t^2|\tilde{p}\rangle \langle \tilde{p}|x_u|\tilde{p}\rangle^2 \tag{33c}$$
$$= \left\{\frac{\partial}{\partial \kappa_{ij}} \frac{\partial}{\partial \kappa_{kl}} \langle \tilde{p}|x_t^2|\tilde{p}\rangle\right\} \langle \tilde{p}|x_u|\tilde{p}\rangle^2$$
$$+ \left\{\frac{\partial}{\partial \kappa_{ij}} \langle \tilde{p}|x_t^2|\tilde{p}\rangle\right\} \left\{2\langle \tilde{p}|x_u|\tilde{p}\rangle \frac{\partial}{\partial \kappa_{kl}} \langle \tilde{p}|x_u|\tilde{p}\rangle\right\}$$
$$+ \left\{\frac{\partial}{\partial \kappa_{kl}} \langle \tilde{p}|x_t^2|\tilde{p}\rangle\right\} \left\{2\langle \tilde{p}|x_u|\tilde{p}\rangle \frac{\partial}{\partial \kappa_{ij}} \langle \tilde{p}|x_u|\tilde{p}\rangle\right\}$$
$$+ \langle \tilde{p}|x_t^2|\tilde{p}\rangle \left\{2 \frac{\partial}{\partial \kappa_{kl}} \langle \tilde{p}|x_u|\tilde{p}\rangle \frac{\partial}{\partial \kappa_{ij}} \langle \tilde{p}|x_u|\tilde{p}\rangle\right\}$$
$$+ \langle \tilde{p}|x_t^2|\tilde{p}\rangle \left\{2\langle \tilde{p}|x_u|\tilde{p}\rangle \frac{\partial}{\partial \kappa_{ij}} \frac{\partial}{\partial \kappa_{kl}} \langle \tilde{p}|x_u|\tilde{p}\rangle\right\}$$

$$\left[\frac{\partial}{\partial \kappa_{ij}} \frac{\partial}{\partial \kappa_{kl}} \langle \tilde{p}|x_t^2|\tilde{p}\rangle \langle \tilde{p}|x_u|\tilde{p}\rangle^2 \right]_{\kappa=0} \tag{33d}$$
$$= \left\{\begin{matrix} +\delta_{ip}\left\{\begin{matrix} \delta_{jk}\langle l|x_t^2|p\rangle - \delta_{jl}\langle k|x_t^2|p\rangle \\ -2\langle j|x_t^2|k\rangle\delta_{lp} + 2\langle j|x_t^2|l\rangle\delta_{kp} \end{matrix}\right\} \\ -\delta_{jp}\left\{\begin{matrix} \delta_{ik}\langle l|x_t^2|p\rangle - \delta_{il}\langle k|x_t^2|p\rangle \\ -2\langle i|x_t^2|k\rangle\delta_{lp} + 2\langle i|x_t^2|l\rangle\delta_{kp} \end{matrix}\right\} \\ +\delta_{kp}\{\delta_{il}\langle j|x_t^2|p\rangle - \delta_{lj}\langle i|x_t^2|p\rangle\} \\ -\delta_{lp}\{\delta_{ik}\langle j|x_t^2|p\rangle - \delta_{jk}\langle i|x_t^2|p\rangle\} \end{matrix}\right\} \langle p|x_u|p\rangle^2$$
$$+ 2\{2\delta_{ip}\langle j|x_t^2|p\rangle - 2\delta_{jp}\langle i|x_t^2|p\rangle\}\langle p|x_u|p\rangle\{2\delta_{kp}\langle l|x_u|p\rangle - 2\delta_{lp}\langle k|x_u|p\rangle\}$$
$$+ 2\{2\delta_{kp}\langle l|x_t^2|p\rangle - 2\delta_{lp}\langle k|x_t^2|p\rangle\}\langle p|x_u|p\rangle\{2\delta_{ip}\langle j|x_u|p\rangle - 2\delta_{jp}\langle i|x_u|p\rangle\}$$
$$+ 2\langle p|x_t^2|p\rangle\{2\delta_{kp}\langle l|x_u|p\rangle - 2\delta_{lp}\langle k|x_u|p\rangle\}\{2\delta_{ip}\langle j|x_u|p\rangle - 2\delta_{jp}\langle i|x_u|p\rangle\}$$
$$+ 2\langle p|x_t^2|p\rangle\langle p|x_u|p\rangle \left\{\begin{matrix} +\delta_{ip}\left\{\begin{matrix} \delta_{jk}\langle l|x_u|p\rangle - \delta_{jl}\langle k|x_u|p\rangle \\ -2\langle j|x_u|k\rangle\delta_{lp} + 2\langle j|x_u|l\rangle\delta_{kp} \end{matrix}\right\} \\ -\delta_{jp}\left\{\begin{matrix} \delta_{ik}\langle l|x_u|p\rangle - \delta_{il}\langle k|x_u|p\rangle \\ -2\langle i|x_u|k\rangle\delta_{lp} + 2\langle i|x_u|l\rangle\delta_{kp} \end{matrix}\right\} \\ +\delta_{kp}\{\delta_{il}\langle j|x_u|p\rangle - \delta_{lj}\langle i|x_u|p\rangle\} \\ -\delta_{lp}\{\delta_{ik}\langle j|x_u|p\rangle - \delta_{jk}\langle i|x_u|p\rangle\} \end{matrix}\right\}$$

**(8/9)**

$$-6\langle \tilde{p}|x_u|\tilde{p}\rangle^2 \langle \tilde{p}|x_t|\tilde{p}\rangle^2 \tag{32}$$

$$\frac{\partial}{\partial \kappa_{ij}} \langle \tilde{p}|x_u|\tilde{p}\rangle^2 \langle \tilde{p}|x_t|\tilde{p}\rangle^2 \qquad [34a]$$

$$= \left\{ 2\langle \tilde{p}|x_u|\tilde{p}\rangle \frac{\partial}{\partial \kappa_{ij}} \langle \tilde{p}|x_u|\tilde{p}\rangle \right\} \langle \tilde{p}|x_t|\tilde{p}\rangle^2$$

$$+ \langle \tilde{p}|x_u|\tilde{p}\rangle^2 \left\{ 2\langle \tilde{p}|x_t|\tilde{p}\rangle \frac{\partial}{\partial \kappa_{ij}} \langle \tilde{p}|x_t|\tilde{p}\rangle \right\}$$

$$\left[ \frac{\partial}{\partial \kappa_{ij}} \langle \tilde{p}|x_u|\tilde{p}\rangle^2 \langle \tilde{p}|x_t|\tilde{p}\rangle^2 \right]_{\boldsymbol{\kappa=0}} \qquad [34b]$$

$$= 2\langle p|x_u|p\rangle \{2\delta_{ip}\langle j|x_u|p\rangle - 2\delta_{jp}\langle i|x_u|p\rangle\} \langle p|x_t|p\rangle^2$$

$$+ 2\langle p|x_u|p\rangle^2 \langle p|x_t|p\rangle \{2\delta_{ip}\langle j|x_t|p\rangle - 2\delta_{jp}\langle i|x_t|p\rangle\}$$

$$\frac{\partial}{\partial \kappa_{ij}} \frac{\partial}{\partial \kappa_{kl}} \langle \tilde{p}|x_u|\tilde{p}\rangle^2 \langle \tilde{p}|x_t|\tilde{p}\rangle^2 \qquad [34c]$$

$$= 2\left\{ \frac{\partial}{\partial \kappa_{kl}} \langle \tilde{p}|x_u|\tilde{p}\rangle \right\} \left\{ \frac{\partial}{\partial \kappa_{ij}} \langle \tilde{p}|x_u|\tilde{p}\rangle \right\} \langle \tilde{p}|x_t|\tilde{p}\rangle^2$$

$$+ 2\langle \tilde{p}|x_u|\tilde{p}\rangle \left\{ \frac{\partial}{\partial \kappa_{ij}} \frac{\partial}{\partial \kappa_{kl}} \langle \tilde{p}|x_u|\tilde{p}\rangle \right\} \langle \tilde{p}|x_t|\tilde{p}\rangle^2$$

$$+ 2\langle \tilde{p}|x_u|\tilde{p}\rangle \left\{ \frac{\partial}{\partial \kappa_{ij}} \langle \tilde{p}|x_u|\tilde{p}\rangle \right\} \left\{ 2\langle \tilde{p}|x_t|\tilde{p}\rangle \frac{\partial}{\partial \kappa_{kl}} \langle \tilde{p}|x_t|\tilde{p}\rangle \right\}$$

$$+ 2\left\{ 2\langle \tilde{p}|x_u|\tilde{p}\rangle \frac{\partial}{\partial \kappa_{kl}} \langle \tilde{p}|x_u|\tilde{p}\rangle \right\} \langle \tilde{p}|x_t|\tilde{p}\rangle \left\{ \frac{\partial}{\partial \kappa_{ij}} \langle \tilde{p}|x_t|\tilde{p}\rangle \right\}$$

$$+ 2\langle \tilde{p}|x_u|\tilde{p}\rangle^2 \left\{ \frac{\partial}{\partial \kappa_{kl}} \langle \tilde{p}|x_t|\tilde{p}\rangle \right\} \left\{ \frac{\partial}{\partial \kappa_{ij}} \langle \tilde{p}|x_t|\tilde{p}\rangle \right\}$$

$$+ 2\langle \tilde{p}|x_u|\tilde{p}\rangle^2 \langle \tilde{p}|x_t|\tilde{p}\rangle \left\{ \frac{\partial}{\partial \kappa_{ij}} \frac{\partial}{\partial \kappa_{kl}} \langle \tilde{p}|x_t|\tilde{p}\rangle \right\}$$

$$\left[\frac{\partial}{\partial \kappa_{ij}}\frac{\partial}{\partial \kappa_{kl}}\langle\tilde{p}|x_u|\tilde{p}\rangle^2\langle\tilde{p}|\hat{x}_t|\tilde{p}\rangle^2\right]_{\kappa=0} \quad [34d]$$
$$= 2\{2\delta_{kp}\langle l|x_u|p\rangle - 2\delta_{lp}\langle k|x_u|p\rangle\}\{2\delta_{ip}\langle j|x_u|p\rangle - 2\delta_{jp}\langle i|x_u|p\rangle\}\langle p|x_t|p\rangle^2$$
$$+ 2\langle p|x_u|p\rangle \left\{\begin{array}{l}+\delta_{ip}\left\{\begin{array}{l}\delta_{jk}\langle l|x_u|p\rangle - \delta_{jl}\langle k|x_u|p\rangle\\-2\langle j|x_u|k\rangle\delta_{lp} + 2\langle j|x_u|l\rangle\delta_{kp}\end{array}\right\}\\-\delta_{jp}\left\{\begin{array}{l}\delta_{ik}\langle l|x_u|p\rangle - \delta_{il}\langle k|x_u|p\rangle\\-2\langle i|x_u|k\rangle\delta_{lp} + 2\langle i|x_u|l\rangle\delta_{kp}\end{array}\right\}\\+\delta_{kp}\{\delta_{il}\langle j|x_u|p\rangle - \delta_{lj}\langle i|x_u|p\rangle\}\\-\delta_{lp}\{\delta_{ik}\langle j|x_u|p\rangle - \delta_{jk}\langle i|x_u|p\rangle\}\end{array}\right\}\langle p|x_t|p\rangle^2$$
$$+ 4\langle p|x_u|p\rangle\{2\delta_{ip}\langle j|x_u|p\rangle - 2\delta_{jp}\langle i|x_u|p\rangle\}\langle p|x_t|p\rangle\{2\delta_{kp}\langle l|x_t|p\rangle$$
$$- 2\delta_{lp}\langle k|x_t|p\rangle\frac{\partial}{\partial \kappa_{kl}}\langle \tilde{p}|x_t|\tilde{p}\rangle\}$$
$$+ 4\langle p|x_u|p\rangle\{2\delta_{kp}\langle l|x_u|p\rangle - 2\delta_{lp}\langle k|x_u|p\rangle\}\langle p|x_t|p\rangle\{2\delta_{ip}\langle j|x_t|p\rangle$$
$$- 2\delta_{jp}\langle i|x_t|p\rangle\}$$
$$+ 2\langle p|x_u|p\rangle^2\{2\delta_{kp}\langle l|x_t|p\rangle - 2\delta_{lp}\langle k|x_t|p\rangle\}\{2\delta_{ip}\langle j|x_t|p\rangle - 2\delta_{jp}\langle i|x_t|p\rangle\}$$
$$+ 2\langle p|x_u|p\rangle^2\langle p|\hat{x}_t|p\rangle \left\{\begin{array}{l}+\delta_{ip}\left\{\begin{array}{l}\delta_{jk}\langle l|x_t|p\rangle - \delta_{jl}\langle k|x_t|p\rangle\\-2\langle j|x_t|k\rangle\delta_{lp} + 2\langle j|x_t|l\rangle\delta_{kp}\end{array}\right\}\\-\delta_{jp}\left\{\begin{array}{l}\delta_{ik}\langle l|x_t|p\rangle - \delta_{il}\langle k|x_t|p\rangle\\-2\langle i|x_t|k\rangle\delta_{lp} + 2\langle i|x_t|l\rangle\delta_{kp}\end{array}\right\}\\+\delta_{kp}\{\delta_{il}\langle j|x_t|p\rangle - \delta_{lj}\langle i|x_t|p\rangle\}\\-\delta_{lp}\{\delta_{ik}\langle j|x_t|p\rangle - \delta_{jk}\langle i|x_t|p\rangle\}\end{array}\right\}$$

**(9/9)**

$$+ 8\langle\tilde{p}|x_t x_u|\tilde{p}\rangle\langle\tilde{p}|x_t|\tilde{p}\rangle\langle\tilde{p}|x_u|\tilde{p}\rangle \quad [33]$$

$$\frac{\partial}{\partial \kappa_{ij}}\langle\tilde{p}|x_t x_u|\tilde{p}\rangle\langle\tilde{p}|x_t|\tilde{p}\rangle\langle\tilde{p}|x_u|\tilde{p}\rangle \quad [35a]$$
$$= \left\{\frac{\partial}{\partial \kappa_{ij}}\langle\tilde{p}|x_t x_u|\tilde{p}\rangle\right\}\langle\tilde{p}|x_t|\tilde{p}\rangle\langle\tilde{p}|x_u|\tilde{p}\rangle$$
$$+ \langle\tilde{p}|x_t x_u|\tilde{p}\rangle\left\{\frac{\partial}{\partial \kappa_{ij}}\langle\tilde{p}|x_t|\tilde{p}\rangle\right\}\langle\tilde{p}|x_u|\tilde{p}\rangle$$
$$+ \langle\tilde{p}|x_t x_u|\tilde{p}\rangle\langle\tilde{p}|x_t|\tilde{p}\rangle\left\{\frac{\partial}{\partial \kappa_{ij}}\langle\tilde{p}|x_u|\tilde{p}\rangle\right\}$$

$$\left[\frac{\partial}{\partial \kappa_{ij}}\langle\tilde{p}|x_t x_u|\tilde{p}\rangle\langle\tilde{p}|x_t|\tilde{p}\rangle\langle\tilde{p}|x_u|\tilde{p}\rangle\right]_{\kappa=0} \quad [35b]$$
$$= \{2\delta_{ip}\langle j|x_t x_u|p\rangle - 2\delta_{jp}\langle i|x_t x_u|p\rangle\}\langle p|x_t|p\rangle\langle p|x_u|p\rangle$$
$$+ \langle p|x_t x_u|p\rangle\{2\delta_{ip}\langle j|x_t|p\rangle - 2\delta_{jp}\langle i|x_t|p\rangle\}\langle p|x_u|p\rangle$$
$$+ \langle p|x_t x_u|p\rangle\langle p|x_t|p\rangle\{2\delta_{ip}\langle j|x_u|p\rangle - 2\delta_{jp}\langle i|x_u|p\rangle\}$$

$$\frac{\partial}{\partial \kappa_{ij}} \frac{\partial}{\partial \kappa_{kl}} \langle \tilde{p}|x_t x_u|\tilde{p}\rangle \langle \tilde{p}|x_t|\tilde{p}\rangle \langle \tilde{p}|x_u|\tilde{p}\rangle \qquad [35c]$$

$$= \left\{ \frac{\partial}{\partial \kappa_{ij}} \frac{\partial}{\partial \kappa_{kl}} \langle \tilde{p}|x_t x_u|\tilde{p}\rangle \right\} \langle \tilde{p}|x_t|\tilde{p}\rangle \langle \tilde{p}|x_u|\tilde{p}\rangle$$

$$+ \left\{ \frac{\partial}{\partial \kappa_{ij}} \langle \tilde{p}|x_t x_u|\tilde{p}\rangle \right\} \left\{ \frac{\partial}{\partial \kappa_{kl}} \langle \tilde{p}|x_t|\tilde{p}\rangle \right\} \langle \tilde{p}|x_u|\tilde{p}\rangle$$

$$+ \left\{ \frac{\partial}{\partial \kappa_{ij}} \langle \tilde{p}|x_t x_u|\tilde{p}\rangle \right\} \langle \tilde{p}|x_t|\tilde{p}\rangle \left\{ \frac{\partial}{\partial \kappa_{kl}} \langle \tilde{p}|x_u|\tilde{p}\rangle \right\}$$

$$+ \left\{ \frac{\partial}{\partial \kappa_{kl}} \langle \tilde{p}|x_t x_u|\tilde{p}\rangle \right\} \left\{ \frac{\partial}{\partial \kappa_{ij}} \langle \tilde{p}|x_t|\tilde{p}\rangle \right\} \langle \tilde{p}|x_u|\tilde{p}\rangle$$

$$+ \langle \tilde{p}|x_t x_u|\tilde{p}\rangle \left\{ \frac{\partial}{\partial \kappa_{ij}} \frac{\partial}{\partial \kappa_{kl}} \langle \tilde{p}|x_t|\tilde{p}\rangle \right\} \langle \tilde{p}|x_u|\tilde{p}\rangle$$

$$+ \langle \tilde{p}|x_t x_u|\tilde{p}\rangle \left\{ \frac{\partial}{\partial \kappa_{ij}} \langle \tilde{p}|x_t|\tilde{p}\rangle \right\} \left\{ \frac{\partial}{\partial \kappa_{kl}} \langle \tilde{p}|x_u|\tilde{p}\rangle \right\}$$

$$+ \left\{ \frac{\partial}{\partial \kappa_{kl}} \langle \tilde{p}|x_t x_u|\tilde{p}\rangle \right\} \langle \tilde{p}|x_t|\tilde{p}\rangle \left\{ \frac{\partial}{\partial \kappa_{ij}} \langle \tilde{p}|x_u|\tilde{p}\rangle \right\}$$

$$+ \langle \tilde{p}|x_t x_u|\tilde{p}\rangle \left\{ \frac{\partial}{\partial \kappa_{kl}} \langle \tilde{p}|x_t|\tilde{p}\rangle \right\} \left\{ \frac{\partial}{\partial \kappa_{ij}} \langle \tilde{p}|x_u|\tilde{p}\rangle \right\}$$

$$+ \langle \tilde{p}|x_t x_u|\tilde{p}\rangle \langle \tilde{p}|x_t|\tilde{p}\rangle \left\{ \frac{\partial}{\partial \kappa_{ij}} \frac{\partial}{\partial \kappa_{kl}} \langle \tilde{p}|x_u|\tilde{p}\rangle \right\}$$

$$\left[\frac{\partial}{\partial \kappa_{ij}}\frac{\partial}{\partial \kappa_{kl}}\langle\tilde{p}|x_t x_u|\tilde{p}\rangle\langle\tilde{p}|x_t|\tilde{p}\rangle\langle\tilde{p}|x_u|\tilde{p}\rangle\right]_{\kappa=0} \quad [35d]$$

$$= \begin{cases} +\delta_{ip}\begin{cases}\delta_{jk}\langle l|x_t x_u|p\rangle - \delta_{jl}\langle k|x_t x_u|p\rangle \\ -2\langle j|x_t x_u|k\rangle\delta_{lp} + 2\langle j|x_t x_u|l\rangle\delta_{kp}\end{cases} \\ -\delta_{jp}\begin{cases}\delta_{ik}\langle l|x_t x_u|p\rangle - \delta_{il}\langle k|x_t x_u|p\rangle \\ -2\langle i|x_t x_u|k\rangle\delta_{lp} + 2\langle i|x_t x_u|l\rangle\delta_{kp}\end{cases} \\ +\delta_{kp}\{\delta_{il}\langle j|x_t x_u|p\rangle - \delta_{lj}\langle i|x_t x_u|p\rangle\} \\ -\delta_{lp}\{\delta_{ik}\langle j|x_t x_u|p\rangle - \delta_{jk}\langle i|x_t x_u|p\rangle\}\end{cases}\langle p|x_t|p\rangle\langle p|x_u|p\rangle$$

$$+\{2\delta_{ip}\langle j|x_t x_u|p\rangle - 2\delta_{jp}\langle i|x_t x_u|p\rangle\}\{2\delta_{kp}\langle l|x_t|p\rangle - 2\delta_{lp}\langle k|x_t|p\rangle\}\langle p|x_u|p\rangle$$

$$+\{2\delta_{ip}\langle j|x_t x_u|p\rangle - 2\delta_{jp}\langle i|x_t x_u|p\rangle\}\langle p|x_t|p\rangle\{2\delta_{kp}\langle l|\hat{x}_u|p\rangle - 2\delta_{lp}\langle k|x_u|p\rangle\}$$

$$+\{2\delta_{kp}\langle l|x_t x_u|p\rangle - 2\delta_{lp}\langle k|x_t x_u|p\rangle\}\{2\delta_{ip}\langle j|x_t|p\rangle - 2\delta_{jp}\langle i|x_t|p\rangle\}\langle p|x_u|p\rangle$$

$$+\langle p|x_t x_u|p\rangle\begin{cases}+\delta_{ip}\begin{cases}\delta_{jk}\langle l|x_t|p\rangle - \delta_{jl}\langle k|x_t|p\rangle \\ -2\langle j|x_t|k\rangle\delta_{lp} + 2\langle j|x_t|l\rangle\delta_{kp}\end{cases} \\ -\delta_{jp}\begin{cases}\delta_{ik}\langle l|x_t|p\rangle - \delta_{il}\langle k|x_t|p\rangle \\ -2\langle i|x_t|k\rangle\delta_{lp} + 2\langle i|x_t|l\rangle\delta_{kp}\end{cases} \\ +\delta_{kp}\{\delta_{il}\langle j|x_t|p\rangle - \delta_{lj}\langle i|x_t|p\rangle\} \\ -\delta_{lp}\{\delta_{ik}\langle j|x_t|p\rangle - \delta_{jk}\langle i|x_t|p\rangle\}\end{cases}\langle p|x_u|p\rangle$$

$$+\langle p|x_t x_u|p\rangle\{2\delta_{ip}\langle j|x_t|p\rangle - 2\delta_{jp}\langle i|x_t|p\rangle\}\{2\delta_{kp}\langle l|x_u|p\rangle - 2\delta_{lp}\langle k|x_u|p\rangle\}$$

$$+\{2\delta_{kp}\langle l|x_t x_u|p\rangle - 2\delta_{lp}\langle k|x_t x_u|p\rangle\}\langle p|x_t|p\rangle\{2\delta_{ip}\langle j|x_u|p\rangle - 2\delta_{jp}\langle i|x_u|p\rangle\}$$

$$+\langle p|x_t x_u|p\rangle\{2\delta_{kp}\langle l|x_t|p\rangle - 2\delta_{lp}\langle k|x_t|p\rangle\}\{2\delta_{ip}\langle j|x_u|p\rangle - 2\delta_{jp}\langle i|x_u|p\rangle\}$$

$$+\langle p|x_t x_u|p\rangle\langle p|x_t|p\rangle\begin{cases}+\delta_{ip}\begin{cases}\delta_{jk}\langle l|x_u|p\rangle - \delta_{jl}\langle k|x_u|p\rangle \\ -2\langle j|x_u|k\rangle\delta_{lp} + 2\langle j|x_u|l\rangle\delta_{kp}\end{cases} \\ -\delta_{jp}\begin{cases}\delta_{ik}\langle l|x_u|p\rangle - \delta_{il}\langle k|x_u|p\rangle \\ -2\langle i|x_u|k\rangle\delta_{lp} + 2\langle i|x_u|l\rangle\delta_{kp}\end{cases} \\ +\delta_{kp}\{\delta_{il}\langle j|x_u|p\rangle - \delta_{lj}\langle i|x_u|p\rangle\} \\ -\delta_{lp}\{\delta_{ik}\langle j|x_u|p\rangle - \delta_{jk}\langle i|x_u|p\rangle\}\end{cases}$$

**Figure Captions**

Figure 14. Structure of $H_{10}$ model systems examined in this study. The nearest-neighbor H-H distance is 1.5 Å.

Figure 2. The isosurfaces of CMOs and LMOs for chain $H_{10}$ ($m = 2$) were plotted using the iso value of 0.06 (obtained at a restricted Hartree-Fock (RHF) level by using STO-3G basis set). Their average fourth moment orbital spreads ($au$) are also presented.

Figure 3. The isosurfaces of CMOs and LMOs for ring $H_{10}$ ($m = 2$) obtained at RHF/STO-3G level.

Figure 4. The isosurfaces of CMOs and LMOs for sheet $H_{10}$ ($m = 2$) obtained at RHF/STO-3G level.

Figure 5. The isosurfaces of CMOs and LMOs for pyramid $H_{10}$ ($m = 2$) obtained at RHF/STO-3G level.

Figure 15. Geometries of the rectangular ground-state ($D_{2h}$) and square transition ($D_{4h}$) structure of cyclobutadiene ($c$-$C_4H_4$) optimized at the CCSD(T)/cc-pVTZ level.[25]

Figure 2. Using the iso value of 0.06, the isosurfaces of CMOs and LMOs of $c$-$C_4H_4$ minimum ($m = 2$) are presented (obtained at HF/cc-pVTZ level). The fourth moment orbital spreads ($au$) are also provided.

Figure 8. Using the iso value of 0.06, the isosurfaces of CMOs and LMOs of $c$-$C_4H_4$ transition state ($m = 2$) are presented (obtained at HF/cc-pVTZ level). The fourth moment orbital spreads ($au$) are also provided.

Figure 16. Frozen core and active orbitals of the rectangular ground-state $c$-$C_4H_4$ in MCSCF calculations.

Figure 17. Localization effect on active space of the equilibrium geometry (contour = 0.06).

Figure 18. Localization effect on active space of the transition state (contour = 0.06).

Figure 19. Optimized geometry of the propargyl radical at GVVPT2/aug-cc-pVTZ level.[32]

Figure 20. The frozen core orbitals of the propargyl radical.

Figure 21. High occupancy canonical MOs in the active space of $1^2B_1$ (ground electronic state).

Figure 22. Localization effect on active space of the ground electronic state (contour = 0.06).

Figure 23. Localization effect on active space of the first excited state (contour = 0.06).

Figure 24. Localization effect on active space of the second excited state (contour = 0.06).

Figure 25. [n]cumulenes.[35,36]

Figure 26. Standard description of $\pi$ bonding is insufficient because extended helices of specific chirality can occur.[36]

Figure 27. Helical localized molecular orbitals.